\documentclass[aps,amsmath,amssymb,showpacs,showkeys]{revtex4}
\usepackage[dvips]{graphicx,color}

\usepackage[utf8]{inputenc}
\usepackage{multirow}
\usepackage{amsmath}
\usepackage{times}
\usepackage{xcolor}
\usepackage{mathtools}
\usepackage[%
  colorlinks=true,
  urlcolor=blue,
  linkcolor=red,
  citecolor=blue
]{hyperref}

\newcommand{\orcid}[1]{\href{https://orcid.org/#1}{\includegraphics[width=8pt]{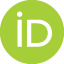}}}

\begin{document}
\title{Thermodynamics and Shadows of GUP-corrected Black Holes with Topological 
Defects in Bumblebee Gravity}

\author{Ronit Karmakar \orcid{0000-0002-9531-7435}}
\email[Email: ]{ronit.karmakar622@gmail.com}

\affiliation{Department of Physics, Dibrugarh University,
Dibrugarh 786004, Assam, India}
\author{Dhruba Jyoti Gogoi \orcid{0000-0002-4776-8506}}
\email[Email: ]{moloydhruba@yahoo.in}

\affiliation{Department of Physics, Dibrugarh University,
Dibrugarh 786004, Assam, India}

\author{Umananda Dev Goswami  \orcid{0000-0003-0012-7549}}
\email[Email: ]{umananda2@gmail.com}

\affiliation{Department of Physics, Dibrugarh University,
Dibrugarh 786004, Assam, India}

\begin{abstract}
In this work we investigate a Schwarzschild-type black hole that is corrected 
by the Generalized Uncertainty Principle (GUP) and possesses topological 
defects within the framework of Bumblebee gravity. Our focus is on the 
thermodynamic characteristics of the black hole, such as temperature, entropy 
and heat capacity, which vary as functions of the horizon radius, and also on 
shadow as an optical feature. Our investigation reveals significant changes 
in the thermodynamic behavior of the black hole due to violations of Lorentz 
symmetry, GUP corrections, and the presence of monopoles. However, the shadow 
of the black hole is unaffected by violations of Lorentz symmetry. In 
addition, we provide a limit on the parameters of Lorentz symmetry violation 
and topological defects based on a classical test involving the precession 
of planetary orbits and the advancement of perihelion in the solar system.

\end{abstract}


\maketitle
\section{Introduction}
\label{sec1}
The General Relativity (GR) has emerged as the most successful theory of 
gravity since its inception and currently it has gained strong support from 
the recent unprecedented observational endeavours with the tremendous 
technological progress. The detection of Gravitational Waves (GWs) 
\cite{1,2,3,4,5} and pictures of the black holes released by the Event Horizon 
Telescope (ETH) group \cite{6,7,8,9,10,11} are two shining milestones in this 
regard. Still, there are enough motivations to look for a better theory of 
gravity due to the shortcomings of GR in many observational and theoretical 
fronts. Two key observational evidences where GR shows its insufficiency to 
provide explanation are the accelerated expansion of the present Universe and 
its missing mass \cite{12,13,14,15,16}. Similarly, if one thinks gravity as a 
fundamental interaction at quantum scale, then GR fails and hence the Standard 
Model (SM) of particle physics could not accommodate gravity in the fold 
of quantum theories along with other three interactions. Though SM explains 
particles and their interaction at microscopic scale, it is unable to deal with 
macroscopic phenomena. The quest to unify GR with SM theories is ongoing with 
some faint light of hope in the form of new directions in the research field. 
Consequently, Quantum Gravity (QG) theories \cite{q1,q2} are being developed 
with the aim of unifying gravity with the quantum field theories and the best 
testing grounds for such a theory, which one can think of right now, is 
probably in the vicinity of black holes. One of the consequences of the Loop 
QG (LQG) \cite{q3,q4} is the possibility of the Lorentz Symmetry Breaking 
(LSB), which can serve as a smoking gun for such a viable quantum theory of 
gravity \cite{18,20}. Thus LSB has a very important role to play in the context
of testing of QG as the energy comparable to Planck scale is needed to test 
such theories of gravity, which is not possible to attain in the present time 
as well as in the foreseeable future. So, as a signature of the QG theories,
the LSB is one of the few options for probing this realm in greater detail. 
As such, recently a lot of attention has been drawn towards the possibility of 
testing the effects of LSB at lower energy scales in different perspectives 
\cite{17,18,19,20,20-1,20-2,20-3,20-4}. 

In the process of circumventing the unsolved problems associated with GR,  
new classes of gravity theories have been developed or already existing 
contemporary ones are gaining importance. One
such class of theories is known as the Modified Theories of Gravity (MTGs),
where the geometric part of GR has been modified in different ways. Some of
the MTGs are the $f(R)$ gravity \cite{51,52,521}, $f(R,T)$ gravity \cite{53},
Rastall gravity \cite{55} etc. Another such class of gravity theories may be
referred to as the Alternative Theories Gravity (ATGs), where the underlying
geometrical structure of spacetime is different from that of GR.
Teleparallel gravity \cite{tp}, Braneworld gravity \cite{bg} etc.\ belong to 
this class of gravity. It needs to be mentioned that an important model of
teleparallel gravity is the $f(Q)$ gravity \cite{54}, which is based on the
symmetric teleparallelism and non-metricity condition. One more class of
gravity theories is usually known as the SM Extension (SME) \cite{sme,sme2}, 
which are basically QG theories wherein the GR is effectively incorporated in 
the SM theories. Thus the Lagrangians of SME models contain the property of the 
LSB \cite{20}. The Bumblebee gravity is the simplest model of SME where the 
Bumblebee vector field acquires a non-vanishing vacuum expectation 
value (VEV) under a suitable potential. The details of this theory can be found
extensively in literature (see e.g.~\cite{17,18,19,20} and references
therein). Motivated by previous works on LSB  in this work we used the
Bumblebee gravity, which carries the characteristic of LSB in the simplest form
\cite{20}. In passing it should be mentioned that these new gravity theories
can explain various phenomena like flat galactic rotation curves \cite{56}
and accelerated expansion of the Universe \cite{12,13} among others.

Spontaneous Symmetry Breaking (SSB) of quantum fields in the early Universe 
may lead to the formation of stable topological defects like 
monopoles \cite{21}. It is stated in Ref.~\cite{18} that these monopoles 
could be the cause of inflation \cite{inf} in the early Universe when there 
were phase transitions of the Universe and hence the Gauge symmetry in the
fields was broken. The effects of the monopole on various properties of a 
Schwarzschild-type black hole was studied in the context of Bumblebee gravity, 
together with the lorentz symmetry parameter in \cite{18}. Recently, Casana 
and his group computed a Schwarzschild-type black hole solution with LSB effect 
and performed three classical tests of GR to give some bounds on the LSB 
parameter \cite{17}. Gogoi and Goswami \cite{20} extensively studied the 
quasinormal modes and sparsity of a Schwarzschild-type black hole corrected by 
the Generalized Uncertainty Principle (GUP) with topological defects in 
Bumblebee gravity. In these works, a combined effect of both LSB and global 
symmetry violation were analysed. It is clear that such topological defects 
have a major implication on the various properties of the black holes. 

The GUP \cite{43,44,45,46,47,48,481,49,50,39,40,40-1,40-2,40-3,40-4,40-5,40-6,40-7,40-8,40-9,40-10, 40-11,40-12,40-13,40-14,40-15,40-16,40-17} 
has been introduced recently in the literature to emphasize
the existence of a minimum length scale at high energy scales. It gives new
insights to theoretical studies of gravity and can provide important novel
intuitions to physicists regarding properties of spacetimes. The Linear and
Quadratic GUP (LQGUP) framework (where GUP with linear and quadratic terms
in momentum are considered) is used in this work, which is inspired
by Refs.~\cite{43,44,45,46,47,48,481}.  Anacleto and collaborators \cite{39,49}
employed a modified mass term $M_{GUP}$ that includes the contributions of
the GUP corrections in the study of the scattering and absorption properties
of a Schwarzschild black hole.  Similarly, L\"{u}tf\"{u}oglu and collaborators
\cite{40} worked out a new type of formalism for incorporating the effects of
GUP in the study of the thermodynamics of a Schwarzschild black hole.
Several other works involving GUP corrections to black holes can be seen in
Refs.~\cite{20,50} with references therein. 

Apart from the black hole physics, GUP has been implemented in different
areas of studies. For example,
in Ref.\ \cite{40-3}, the authors discussed the emergent Universe 
model with GUP and concluded that GUP-based cosmology can replicate the 
emergent Universe scenario comprehensively. In Ref.\ \cite{40-4}, the authors 
discussed four astrophysical phenomena and the influence of GUP on such 
phenomena. Similarly, in Ref.\ \cite{40-8}, the authors have demonstrated the 
equivalence principle with application of linear and quadratic GUP in 
obtaining analogy to Liouville theorem and various other fields like density 
of states, black body radiation among others. Further, in Ref.\ \cite{40-9}, 
the authors calculated the Shapiro time delay, geodesic precession and 
gravitational redshift for the Schwarzschild-GUP metric and constrained the 
GUP parameter $\beta$ using the solar system experiments.
 However, the observational constraints on the GUP parameters is one of 
the current focus points of concern. In this respect we would like to mention 
that the Quasinormal Modes (QNMs) of oscillations \cite{20} of black holes can 
provide a strong ground regarding this. Future GW detectors like LISA would 
be able to detect QNMs, which will help to constrain GUP parameters associated 
with various models. Another possible way of constraining them is the use of 
observational data of black hole shadow \cite{shadow_jusufi}, made available 
by EHT recently. Mainly M87* and SgrA* black holes' shadows data have been 
utilised in literature to constrain various parameters of the related theory 
\cite{shadow_jusufi}. We are very hopeful that GUP parameters can also be well 
constrained with the shadow data.

Black Hole is a region of spacetime where the curvature of spacetime 
is so huge that it creates a boundary of no return. Within this region, 
if anything enters, then it cannot leave the region due to the enormous 
curvature of spacetime or the immense gravity of the black hole. Black holes 
show thermodynamic properties analogous to a thermodynamic system and four 
laws of black hole thermodynamics have been proposed in this regard 
\cite{22,23,24,25,26,27}. Classically, we can imagine a black hole to only 
absorb radiation and matter, but quantum mechanical treatment proves that 
black hole can also emit radiation. This property establishes that black holes 
are thermodynamic systems. Many recent studies have been carried out to 
understand the black hole thermodynamics in terms of temperature, radiation 
sparsity, entropy, area quantization and various other related properties 
\cite{28,29,30,31,32,33,35,36,37,38,39,40,41,42,dj}. 

Shadow of black holes has been studied extensively in literature
\cite{1s,shnew02,shnew01,2s,3s,4s,5s,6s,7s,8s,9s,10s,11s,12s,13s,14s,15s,
16s,17s,18s,19s,102-1,102-2,102-3,102-4,102-5,102-6,102-7,102-8,102-9,102-10,102-11,
102-12,102-13,102-14,102-15,102-16} and has gained importance after the released photographs of the black 
holes inrecent times. Shadow of a black is its apparent shape when it is illuminated 
by a background light source and is an important phenomenological feature. 
The specific feature of a shadow depends on the physical properties of the 
related black hole \cite{3s}. Thus shadows can be used to extract information 
on the physical properities of the associated black holes. Moreover, shadows 
can be used to differentiate between theories of gravity as they are specific 
to black holes' physical properties \cite{3s}. In recent times, many works 
have been focussed on this aspect. K.\ Jusufi \cite{1s} studied the shadow and 
QNMs of a black hole surrounded by dark matter and 
established a relation between the real part of the QNMs and the shadow radius 
of the black hole. Circular shadow radius was obtained in this case. A similar 
kind of work was performed previously in Ref.\ \cite{15s} as well. In Ref.\ 
\cite{shnew01} black hole shadow in symmergent gravity, that is $R+R^2$ 
gravity in vacuum as well as in plasma background has been studied. 
Authors in Ref.\ \cite{102-2} considered the Einstein-{\AE}ther 
gravity and investigated various properties like photon sphere radius, 
potential and angular size of shadow, while comparing with observationally 
available M87* shadow data. They also imposed constraints and the upper bound 
on the model parameters. The usefulness of shadow observations of supermassive 
black holes to test the Kerr metric has been discussed in Ref.\ \cite{102-3}. 
In Ref.\ \cite{102-4}, authors calculated the shadow of Loop quantum gravity 
inspired black hole solution and by employing the Newmann-Janis algorithm, 
they calculated the rotating counterpart solution. They also studied the 
super-radiance of the rotating solution and found that with increasing mass, 
the difference from the general Kerr case begins to decrease. Authors in 
Ref.\ \cite{102-5} by considering the Johannsen and Psaltis metric, performed 
a test of the No-hair theorem using the EHT data. They reported that the hairy 
Kerr black hole solution plays the role of alternative compact object instead 
of Kerr black hole. Authors in Ref.\ \cite{102-6} employ the shadow of M87* and 
Sgr A* to constrain LSB-impacted Shwarzschild solution and have provided 
bounds on the model parameters from observations. Another work \cite{102-7} 
shows the possibility of existence of extra spatial dimension from the shadow 
of Sgr A*. One more interesting work \cite{102-8} utilised two Loop quantum 
gravity inspired rotating black hole solutions and also observational data of 
shadows, to constrain the model deviation parameter so that the model result 
fits with the observations. It was also concluded that the bounds were more 
effective for the Sgr A* case than that of M87*. Inspired by the works 
mentioned above, we intend to study the behaviour of the black hole shadow 
formed by the black hole that we shall consider. 

Ref.~\cite{17} discusses some classical tests of GR which are also satisfied 
by any theory of gravity. Through this analysis, they established some upper 
bounds on the LSB parameter. We follow the suit and perform the perihelion 
analysis for our black hole metric. Perihelion precession of planets around 
the sun means that on completing one orbit of revolution round the sun, the 
position of the planet advances through a small linear distance from its 
previous position, much similar to the concept of pitch of a rotating screw. 
This analysis provides an upper bound on the value of the LSB term, the 
topological defect or monopole term and the GUP parameters as we will see.

 Although it is already clear, at this point we wish to emphasize the fact
that LSB can possibly lead beyond SM physics. Many experiments 
have tried to establish beyond SM physics by considering LSB effects 
\cite{ref1,ref2}. Also LSB is a feature not exclusive to Bumblebee gravity but 
also occurs in theories like string theory and non-commutative geometries. 
As mentioned earlier, it is established that when symmetry breaks spontaneously 
in such theories, many topological defects, like domain walls, cosmic strings 
or monopole solutions etc.\ can occur. It should be noted that by the Kibble 
mechanism \cite{ref3}, there is a possibility that such topological defects 
arise in the early Universe during some phase transition stages and since 
these isolated defects are stable, they can exist in present times as well. 
These relics could potentially play a role in cosmological phenomena like 
large structure formation and in the  behaviours of astrophysical compact 
objects, such as neutron stars, black holes etc. It is the reason that 
motivates us to consider the global monopole into our model and study its 
influence on various properties of the black hole metric. Moreover, as 
discussed earlier the GUP, which is the modified dispersion relation based on 
minimum length or maximum momentum, is also endowed with the LSB behaviour. 
However, in our work, we consider that the LSB is caused by the Bumblebee 
field only, while the GUP will be used to modify the mass term of the 
considered black hole. 
It is noteworthy to mention that the GUP together with the Bumblebee scenario 
has been already considered in literature \cite{ref4}. Hence, it is expected 
that for a complete understanding of a black hole system in the Bumblebee 
gravitational framework, inclusion of GUP and global monopole seems to be 
obligatory.

We organise the rest of the paper as follows. In section \ref{sec2}, for the 
completeness we discuss the basic mathematical framework of Bumblebee gravity 
along with implementation of LQGUP. In the next section \ref{sec3}, 
the expressions for various thermodynamic properties associated with the black 
hole metric like temperature, entropy, heat capacity ete.\ have been derived 
and then discuss the numerical results of these properties for different 
parameters of the model considered in this work. In section \ref{sec4}, we 
study the variation of the shadow radius with various parameters of the model. 
We derive a classical upper bound on the model parameters using observational 
perihelion precession data of various planets as well as a known comet in 
section \ref{sec5}. In the last section, we present the concluding remarks of 
the work and also discuss some future possibilities related to this work. Here
we adopt a unit system, where $c = \hbar = G = 1$.

\section{Bumblebee gravity and LQGUP corrections}
\label{sec2}
The basic derivation of the metric form that we have used in our work is 
adopted from the Ref.~\cite{20}, where detailed derivation can be inferred. 
Here we mention only some steps in the derivation of the black hole solution.
The Lagrangian density corresponding to the Bumblebee field coupled to gravity 
with topological defects can be written as \cite{20}
\begin{equation}
\mathcal{L}=\sqrt{-g} \bigg[\frac{1}{2\kappa} (R-2\Lambda) +\frac{\xi}{2\kappa} \,\mathcal{B}^{\mu}\mathcal{B}_{\mu}R_{\mu\nu}-\frac{1}{4}\,\mathcal{B}^2_{\mu\nu}-V\big(\mathcal{B}^{\mu}\mathcal{B}_{\mu}\pm b^2\big)\bigg]+\mathcal{L}_{M},
\label{Eq:1}
\end{equation} 
where $\Lambda$ represents the cosmological constant and $\kappa=8\pi$. Here 
the Bumblebee field is represented as $\mathcal{B}_{\mu}$, the field strength 
tensor $\mathcal{B}_{\mu\nu}=\partial_{\mu}\mathcal{B}_{\nu}-\partial_{\nu} 
\mathcal{B}_{\mu}$. The potential associated with the Bumblebee field is 
$V(\mathcal{B}^{\mu}\mathcal{B}_{\mu}\pm b^2)$ with $b^2$ as a positive 
parameter responsible for causing SSB. $\xi$ represents the coupling term and 
$\mathcal{L}_M$ is the Lagrangian density due to the global monopole. 
The field equation for the theory is derived from the Lagrangian \eqref{Eq:1} 
by varying its action with respect to $g_{\mu\nu}$ and is given by
\begin{equation}
G_{\mu\nu}=\kappa \left(T^{\mathcal{B}}_{\mu\nu}+T^{M}_{\mu\nu}\right),
\label{Eq:2}
\end{equation}
where $T^{\mathcal{B}}_{\mu\nu}$ is the energy-momentum tensor that 
depends on the Bumblebee field whose form can be written as \cite{20}
\begin{align}
T_{\mu\nu}^{\mathcal{B}}\equiv & -\mathcal{B}_{\mu\sigma}\mathcal{B}_{\phantom{\sigma}\nu}^{\sigma}-\frac{1}{4}g_{\mu\nu}\mathcal{B}_{\alpha\beta}^{2}-g_{\mu\nu}V\!\left(\mathcal{B}^{\mu}\mathcal{B}_{\mu}\right)+4V^{\prime}\mathcal{B}_{\mu}\mathcal{B}_{\nu}\nonumber \\
 & +\frac{\xi}{\kappa}\left(\frac{1}{2}g_{\mu\nu}\mathcal{B}^{\alpha}\mathcal{B}^{\beta}R_{\alpha\beta}-\mathcal{B}_{\nu}\mathcal{B}^{\alpha}R_{\alpha\mu}-\mathcal{B}_{\mu}\mathcal{B}^{\alpha}R_{\alpha\nu}\right)\nonumber \\
 & +\frac{\xi}{\kappa}\left(\frac{1}{2}\nabla_{\alpha}\nabla_{\mu}\!\left(\mathcal{B}^{\alpha}\mathcal{B}_{\nu}\right)+\frac{1}{2}\nabla_{\alpha}\nabla_{\nu}\!\left(\mathcal{B}_{\mu}\mathcal{B}^{\alpha}\right)\right)\nonumber \\
 & +\frac{\xi}{\kappa}\left(-\frac{1}{2}\nabla^{\lambda}\nabla_{\lambda}\!\left(\mathcal{B}_{\mu}\mathcal{B}_{\nu}\right)-\frac{1}{2}g_{\mu\nu}\nabla_{\alpha}\nabla_{\beta}\!\left(\mathcal{B}^{\alpha}\mathcal{B}^{\beta}\right)\right),\label{T_mn^B}
\end{align}
where as usual the prime is used to denotes the derivative with respect to the
field $\mathcal{B}_{\mu}$. $T^M_{\mu\nu}$ represents the energy-momentum 
tensor for the global monopole contribution and is given by \cite{20}
\begin{equation}
T^{M \nu}_{\mu}= \text{diag}\bigg(\frac{\eta^2}{r^2},\frac{\eta^2}{r^2},0,0\bigg),
\label{Eq:3}
\end{equation}
where $\eta$ is the global monopole parameter. 
Further, the action of the Lagrangian \eqref{Eq:1} can be varied with respect 
to the Bumblebee field to obtain another field equation as
\begin{equation}
\nabla \mathcal{B}_{\mu\nu} = \mathcal{J}^{\mathcal{B}}_\nu + \mathcal{J}^{M}_\nu,
\label{Eq:4}
\end{equation}
where $\mathcal{J}^{\mathcal{B}}_\nu $ and $\mathcal{J}^{M}_\nu$ are 
respectively the self interacting current and the source current associated 
with the Bumblebee field \cite{20}. The potential associated with the 
Lagrangian \eqref{Eq:1} generates a non-vanishing vacuum expectation value 
for the field $\mathcal{B}_{\mu}$ such that
\begin{equation}
\mathcal{B}_{\mu}\mathcal{B}^{\mu} \pm b^2 =0.
\label{Eq:5}
\end{equation}
The above equation yields a solution $\langle\mathcal{B}_{\mu}\rangle=b_{\mu}$,
in which $b_{\mu}$ represents a vector field which spontaneously violates the 
Lorentz symmetry. 

Now proceeding with the Birkhoff metric as standard ansatz,
\begin{equation}
g_{\mu\nu}=\text{diag}\left(-\,e^{2\gamma},e^{2\rho},r^2,r^2 \sin^2 \theta\right),
\label{Eq:6}
\end{equation}
where $\gamma$ and $\rho$ are some arbitrary functions of $r$ and considering 
$\mathcal{B}_{\mu}=b_{\mu}$, i.e.\ taking the Bumblebee field at its vacuum 
expectation value, we solve the field Eq.~\eqref{Eq:2} in vacuum, which leads
the solutions as \cite{20}
\begin{align}
e^{2\rho} & = (1+\lambda)\Big(1+\eta^2 -\frac{\rho_{0}}{r}\Big)^{-1}\!\!\!,
\label{Eq:7}\\[5pt]
e^{2\gamma} & = 1+\eta^2 -\frac{\rho_{0}}{r}.
\label{Eq:8}
\end{align}
Thus the spherically symmetric solution admitting the LSB and global monopole 
can be stated as 
\begin{equation}
ds^2=-\bigg(1-\mu-\frac{2M}{r}\bigg)dt^2 +(1+\lambda)\bigg(1-\mu-\frac{2M}{r}\bigg)^{-1} dr^2 +r^2 d\theta^2+r^2 \sin^2 \theta d\phi^2,
\label{Eq:9}
\end{equation}
where we used $\rho_0=2M$, $M$ being the mass of the black hole, and 
$\mu=-\,\eta^2$ represents the global monopole term. This metric is further 
modified using the LQGUP corrections, incorporating the concept of minimum 
length scale. The final form of the LQGUP-modified metric with Lorentz symmetry 
violation and global monopole is given by \cite{20}
\begin{equation}
ds^2=-\bigg(1-\mu-\frac{2M_{GUP}}{r}\bigg)dt^2 +(1+\lambda)\bigg(1-\mu-\frac{2M_{GUP}}{r}\bigg)^{-1} dr^2 +r^2 d\theta^2+r^2 \sin^2 \theta d\phi^2,
\label{Eq:10}
\end{equation} 
where $M_{GUP}=M\left(1-\alpha (1-\mu)/4M + \beta (1-\mu)^2/8M^2\right)$ is 
the LQGUP corrected mass of the black hole, and $\alpha$ and $\beta$ are the 
GUP parameters. Using this Eq.~\eqref{Eq:10} we proceed further to determine 
various properties of the black hole in the following section. At this point 
it is noteworthy that the horizon radius of the black hole can be computed from 
the above metric using the condition on the metric function as 
\begin{equation}
f(r)|_{r\,=\,r_H}=0,
\label{Eq:11}
\end{equation}
which gives
\begin{equation}
r_H=\frac{2M_{GUP}}{1-\mu}=\frac{2M}{1-\mu} \Big(1-\frac{\alpha (1-\mu)}{4M}+\frac{\beta (1-\mu)^2}{8M^2}\Big).
\label{Eq:12}
\end{equation}
It is seen that the horizon radius of the black hole depends on the global 
monopole term $\mu$ as expected, but it is independent of LSB parameter 
$\lambda$. 

\section{Thermodynamic features of the black hole}
\label{sec3}
The inclusion of GUP corrections along with LSB and global monopole can lead 
to new insights in the thermodynamic perspective of the black holes. At the 
Planck scale of energy, it becomes necessary to incorporate the concept of a 
minimum length. The metric \eqref{Eq:10} incorporates the effects of such a 
minimum length and also the effects of LSB and global monopole. Thermodynamics 
of black holes have been studied in detail in literature as mentioned earlier 
and here we briefly present the calculations of various thermodynamic features 
of the Schwarzschild-like black holes and their variations with various parameters of the formalism.

 To this end, it is important to mention that the Bumblebee gravity 
theory that we consider in our work is a vector-tensor theory of gravity. 
Therefore the first law of black hole thermodynamics for this theory of gravity
together with topological defects can be stated as
\begin{equation}
dE = T dS + \mathcal{A}\, d\mu, 
\label{mt}
\end{equation}
where $E$ is the total energy associated with the stationary spacetime, 
$\mu $ is the magnetic monopole term and $\mathcal{A}$ is the monopole 
potential. It is established in Ref.\ \cite{ref5} that for a simple class of 
vector-tensor theory like Bumblebee theory, the first law of black hole 
thermodynamics need to be modified, and instead of the black hole mass term, 
we have to make use of the Komar mass ($M_K$) of the black hole 
\cite{ref6}. Komar mass of a black hole spacetime is calculated by using the 
relation \cite{ref6}: $M_K=\frac{r^2}{2}\sqrt{\frac{g(r)}{f(r)}}f'(r)$, 
where $f(r)$ and $g(r)$ are the time and radial components respectively of a 
metric and for our metric \eqref{Eq:10}, we find the expression,
\begin{equation}
 M_K=\frac{M \Big(\frac{\beta  (1-\mu )^2}{8 M^2}-\frac{\alpha  (1-\mu )}{4 M}+1\Big)}{\sqrt{1+\lambda }}.
\label{komar}
\end{equation}
Hence, we can rewrite the first law of black hole thermodynamics as 
\begin{equation}
    dM_K=TdS+\mathcal{A}\, d\mu.
    \label{mtl}
\end{equation}
Now for examining the validity of the first law of black hole thermodynamics, 
we calculate the temperature of the black hole from this thermodynamic 
relation \eqref{mtl} at constant $\mu$ as
\begin{equation}
T=\frac{d M_K}{dS}\Big|_{\mu}.
\end{equation}
Substituting the expression of the Komar mass $M_K$ from Eq.\ \eqref{komar} 
in above equation, we find the temperature of the black hole as given by
\begin{equation}
    T=\frac{8M^2 - 2M\alpha (1-\mu)+\beta(1-\mu)^2}{16\pi r_H^2 \sqrt{1+\lambda}}.
    \label{t1}
\end{equation}

Again, the temperature associated with a black hole can also be found out 
from relation \cite{18},
\begin{equation}
T_{BH}=\frac{1}{4\pi}\frac{f'(r)}{\sqrt{f(r) g(r)}}\Big|_{r_H},
\label{Eq:13}
\end{equation}
which leads to the thermodynamic temperature of the black hole defined by the 
metric \eqref{Eq:10} as
\begin{equation}
T_{BH}=\frac{8M^2 -2M \alpha(1-\mu)+\beta(1-\mu)^2}{16M \pi r_H^2 \sqrt{1+\lambda}}.
\label{Eq:14}
\end{equation}
 The exact form of the two expressions \eqref{t1} and \eqref{Eq:14} of the 
black hole temperature justifies that the first law of black hole 
thermodynamics holds in our case of Bumblebee gravity with topological 
defects. It is seen from the expressions that both monopole and LSB parameters
have decreasing effect on the black hole temperature.

For the completeness we can derive the expression for the monopole potential 
from Eq.\ \eqref{mtl} as follows:
\begin{equation}
    \mathcal{A}=\frac{dM_K}{d\mu}\Big|_{S}=-\frac{\beta  (1-\mu )}{4 \sqrt{\lambda +1} \Big(\frac{M (\alpha +2 r)}{\sqrt{(\alpha +2 r)^2-8 \beta }}+M\Big)}.
\end{equation}

As already mentioned, since the black hole mass $M$ fails to satisfy the first 
law of black hole thermodynamics, we implemented the Komar mass in its place 
which is found to be validating the first law. Hence, in the rest of our 
investigation of thermodynamic quantities, we shall use the Komar mass as 
the energy term in the first law.

\begin{figure}[h!]
\includegraphics[scale=0.32]{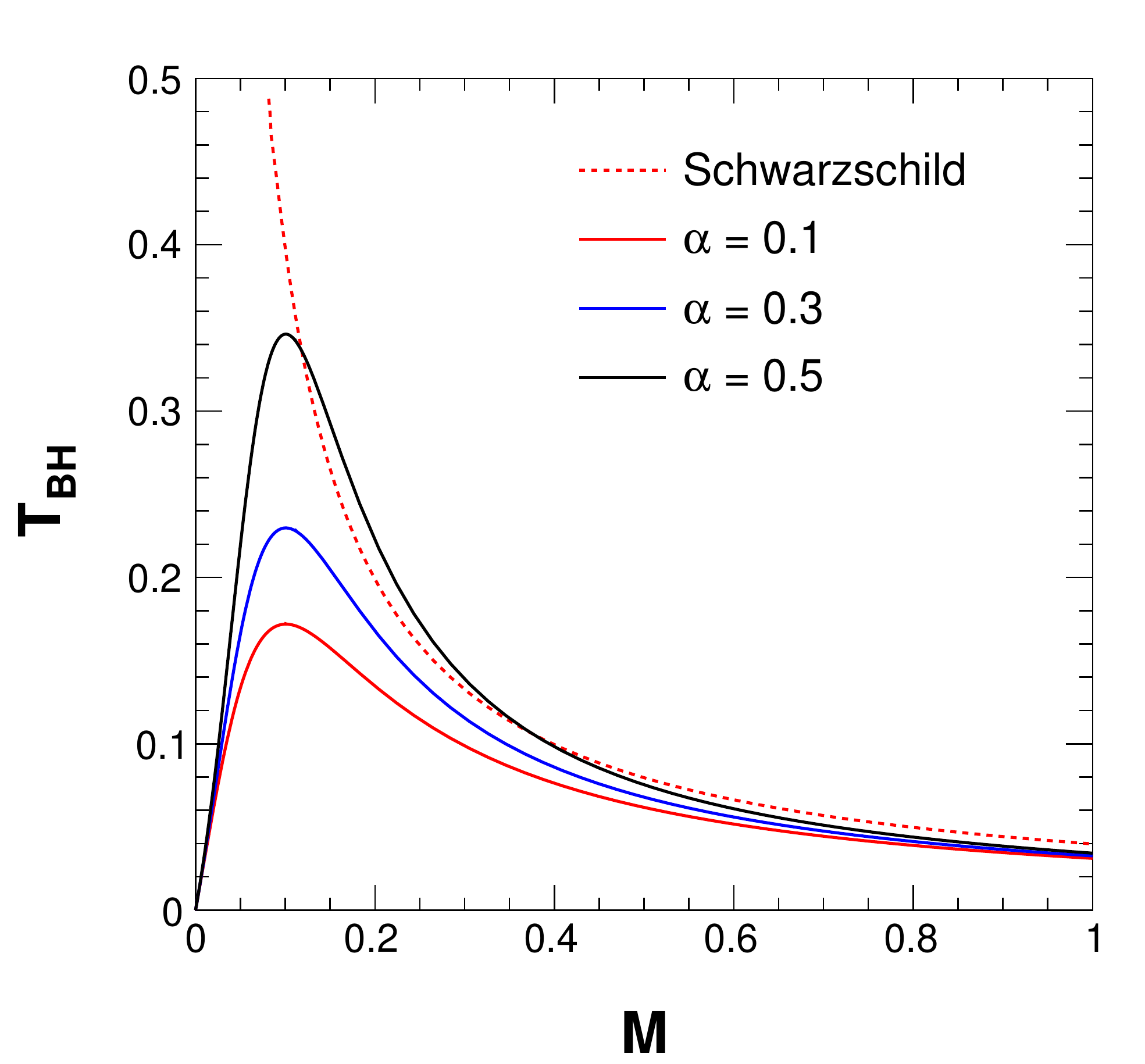}\hspace{0.5cm}
\includegraphics[scale=0.32]{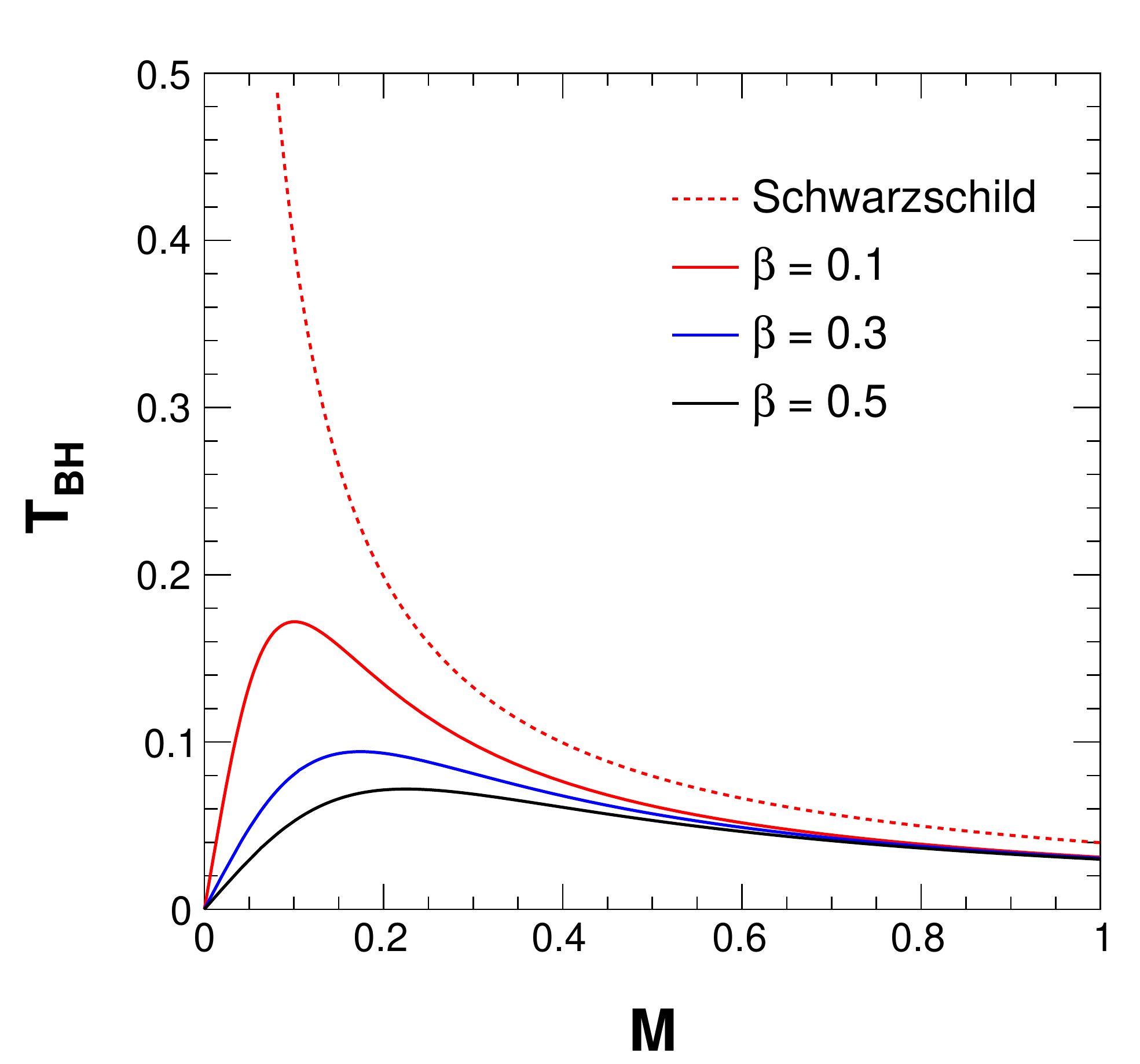}\vspace{0.2cm}
\includegraphics[scale=0.32]{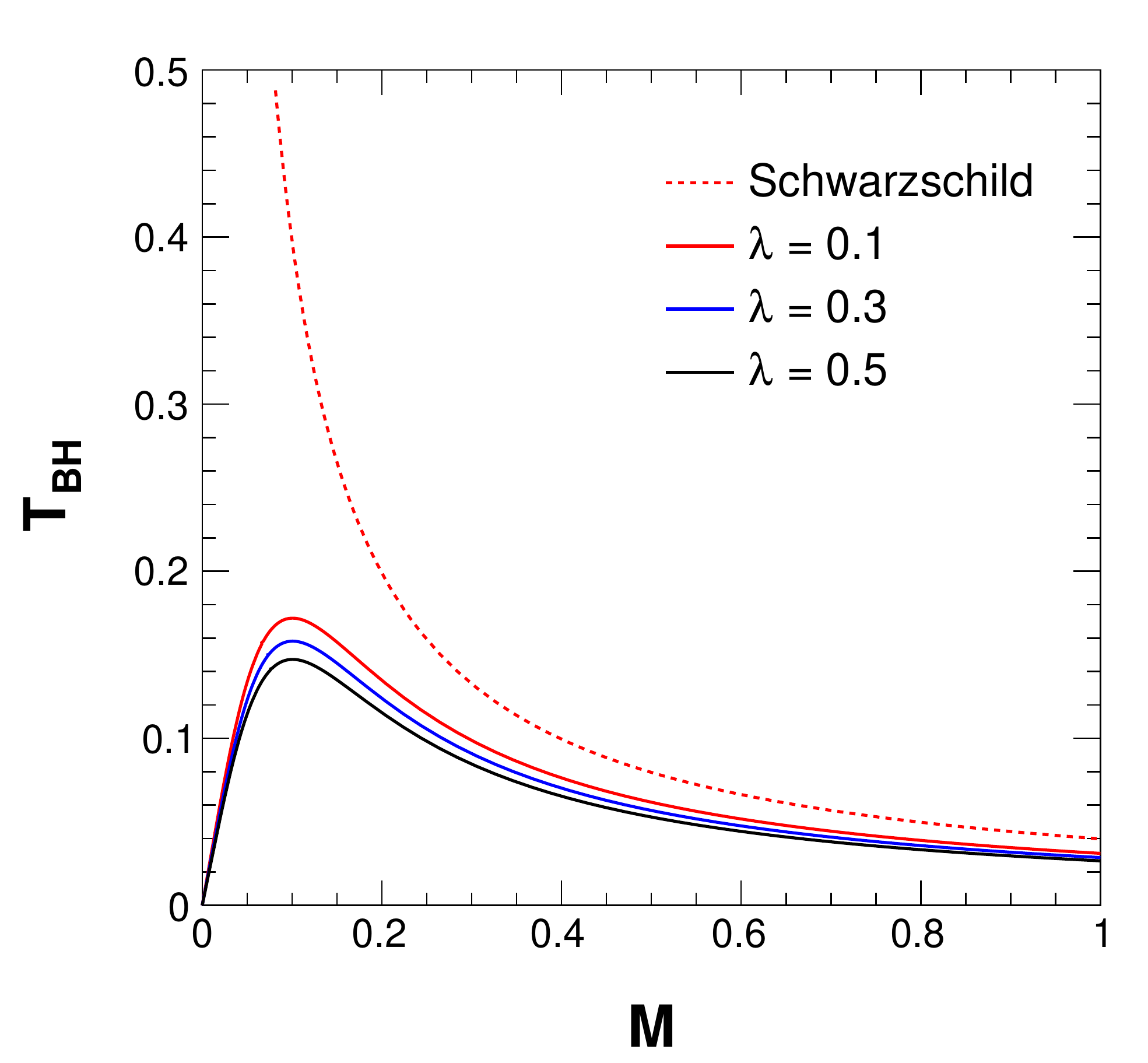}\hspace{0.5cm}
\includegraphics[scale=0.32]{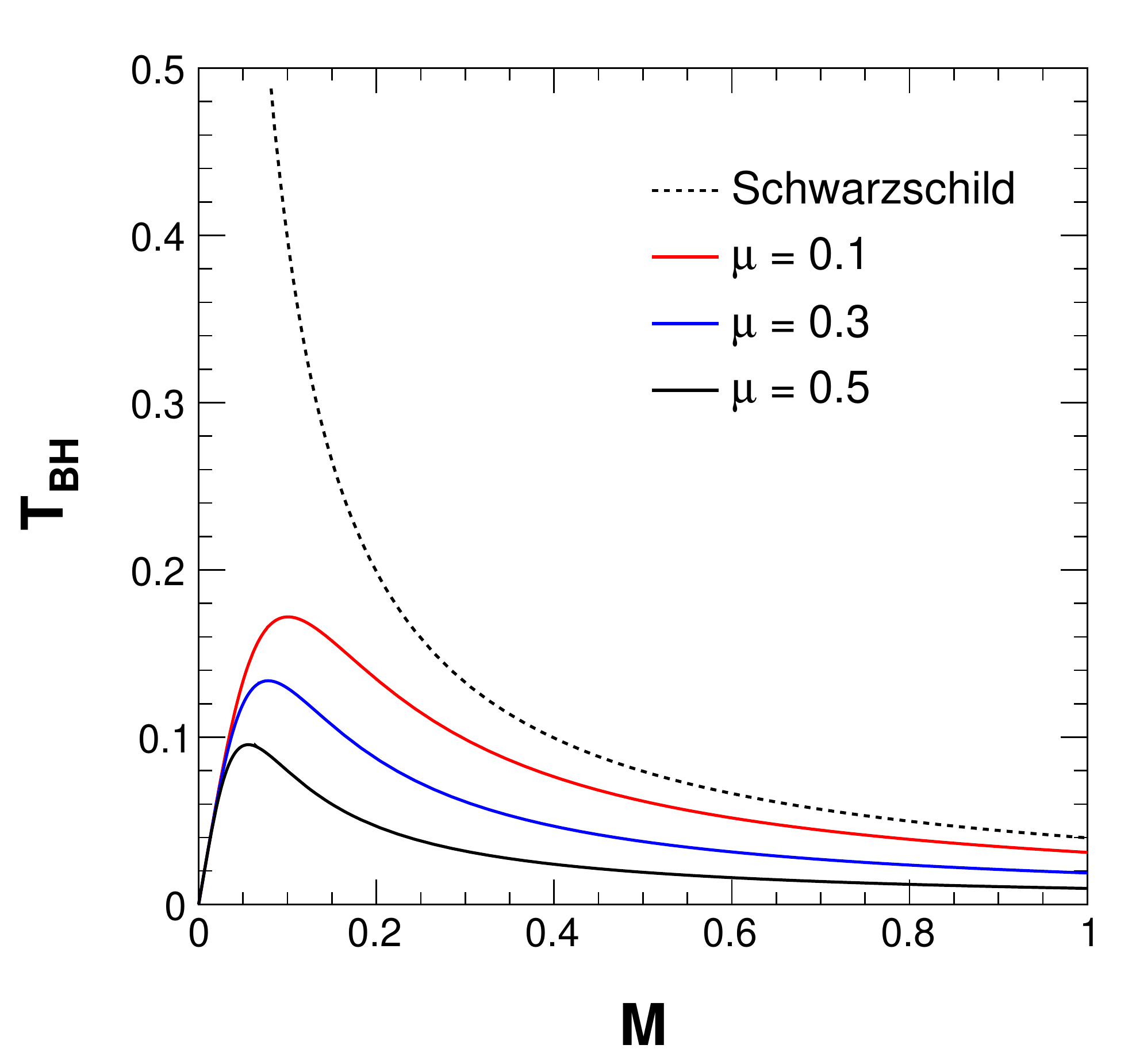}
\vspace{-0.5mm}
\caption{ Variation of temperature of the black hole \eqref{Eq:10} with
respect to mass for different values of $\alpha$, $\beta$, $\lambda$ and 
$\mu$ as obtained from Eq.~\eqref{Eq:14}. We use $\mu =\beta =\lambda=0.1$ for 
the first plot, $\mu = \alpha = \lambda = 0.1$ for the second plot, 
$\mu=\alpha=\beta=0.1$ for the third plot and $\lambda=\alpha=\beta=0.1$ 
for the fourth plot respectively. The dotted line in each plot denotes the 
Schwarzschild case.}
\label{Fig01}
\end{figure}

The variation of temperature of the black hole with mass $M$ for 
different values of the parameters of the theory as given by the expression
\eqref{Eq:14} is shown in Fig.~\ref{Fig01}.
 As seen in the plots, there is a little variation of temperature with 
change in the LSB parameter $\lambda$ (third plot). Whereas there is 
significant variations in temperature for the GUP parameters $\alpha$ and  
$\beta$, and the magnetic monopole parameter $\mu$. It is clear that with 
higher values of the GUP parameter $\alpha$, the peak value of temperature 
increases. The opposite situation can be seen for the other parameter $\beta$. 
The same decreasing behaviour is seen for $\mu$ and $\lambda$ as expected from
the expression. It is to be noted that the temperature is positive, though it 
becomes smaller with increasing $M$. Moreover, for the small values of $M$, 
depending on the model parameter the value of temperature and its pattern are 
substantially different from the Schwarzschild case.

\begin{figure}[h!]
\includegraphics[scale=0.32]{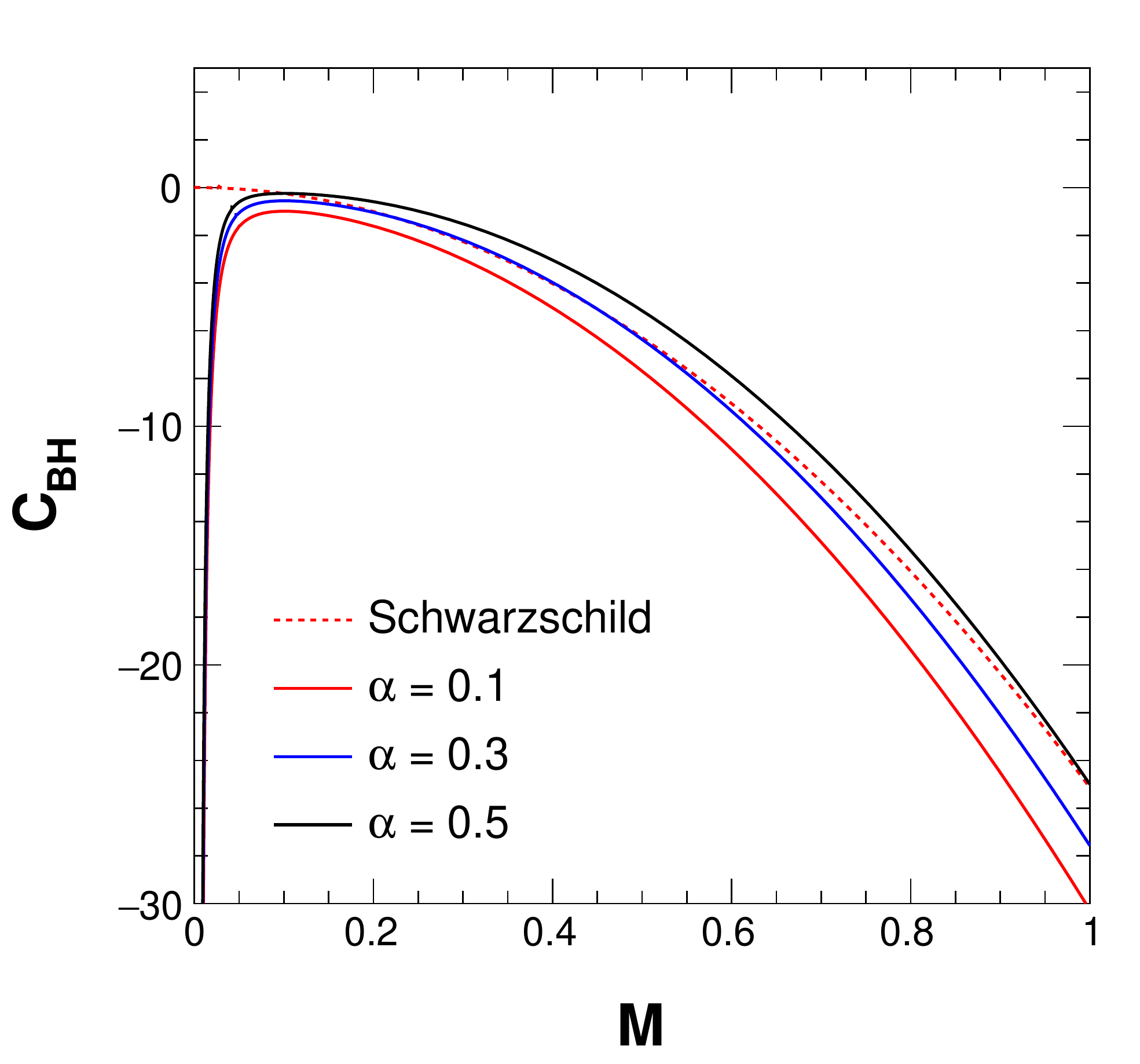}\hspace{0.5cm}
\includegraphics[scale=0.32]{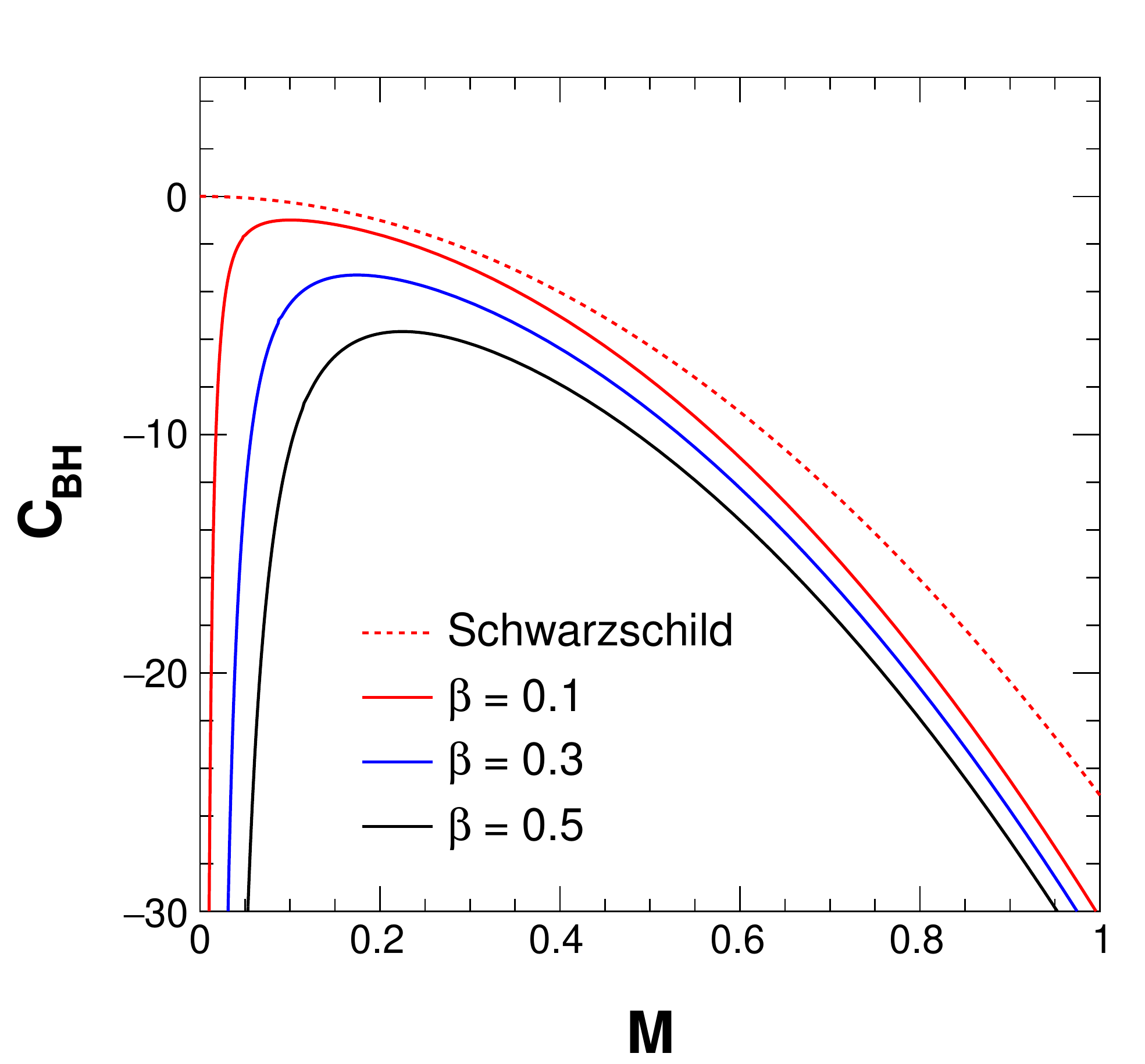}\vspace{0.2cm}
\includegraphics[scale=0.32]{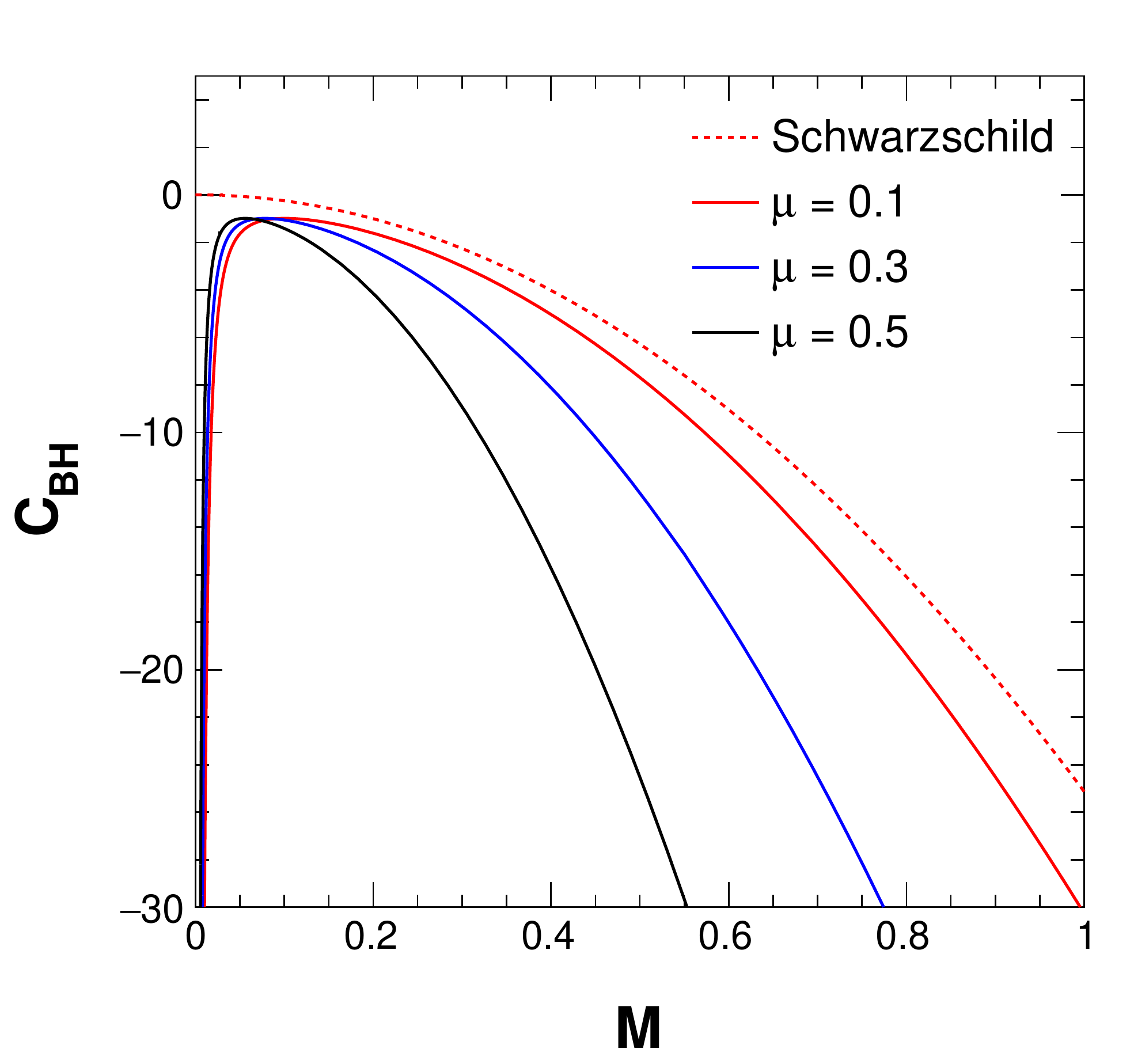}
\caption{ Variation of heat capacity with respect to mass for different 
values of $\alpha$, $\beta$ and $\mu$. We use $\beta= \mu = 0.1$ for the 
first plot, $\alpha = \mu = 0.1$ for the second and $\alpha = \beta = 0.1$ 
for the third plot respectively.}
\label{Fig02}
\end{figure}

 Next, we compute the heat capacity of the black hole from a relation 
obtained by using Eq.\ \eqref{mtl}, considering $\mu$ as constant, given by
\begin{equation}
  C_{BH}=\frac{d M_K}{dT}\Big|_{\mu}=T\,\frac{dS}{dT}. 
\end{equation}
This leads to the expression of heat capacity as
\begin{equation}
 C_{BH}=-2\pi r_H^2= -\frac{\pi  \Big(\beta  (\mu -1)^2+8 M^2+2 \alpha  (\mu -1) M\Big)^2}{8 (\mu -1)^2 M^2}.
\label{Eq:15}
\end{equation}
We plot the variations of $C_{BH}$ with respect to the black hole mass 
for different associated parameters of the model in Fig.\ \ref{Fig02}. The 
first plot is for various values of $\alpha$, second one for $\beta$ and 
third one for $\mu$ . From the plots, it is observed that the heat capacity 
for the black hole always remains negative, signalling that the black hole is 
thermodynamically unstable. It can be seen that heat capacity is independent 
of LSB parameter $\lambda$. It is also clear from the plots of the figure that 
$C_{BH}$ decreases with increasing $M$ and as $M$ decreases towards zero, 
$C_{BH}$ becomes more and more negative, specifying the different behaviour of
the black hole from the Schwarzschild one in this respect. The fact is that 
when a black hole emits more than it absorbs, then there is more probability 
of evaporation of the black hole, signifying instability.

It is to be noted that in this case, there is no possibility of remnant 
formation mathematically as clear from Eq.\ \eqref{Eq:15}. The criteria for 
remnant formation states that $C_{BH}=0$ should be solved to find out the 
remnant radius, which when employed into the temperature expression 
\eqref{Eq:14}, gives the remnant temperature \cite{40}. The study of remnant 
formation is important in the sense that it provides an alternative to the 
theory of complete evaporation of the black hole. Once a remnant forms, it 
does not emit any radiation which may make it difficult to observe them 
directly. 

\begin{figure}[h!]
\includegraphics[scale=0.32]{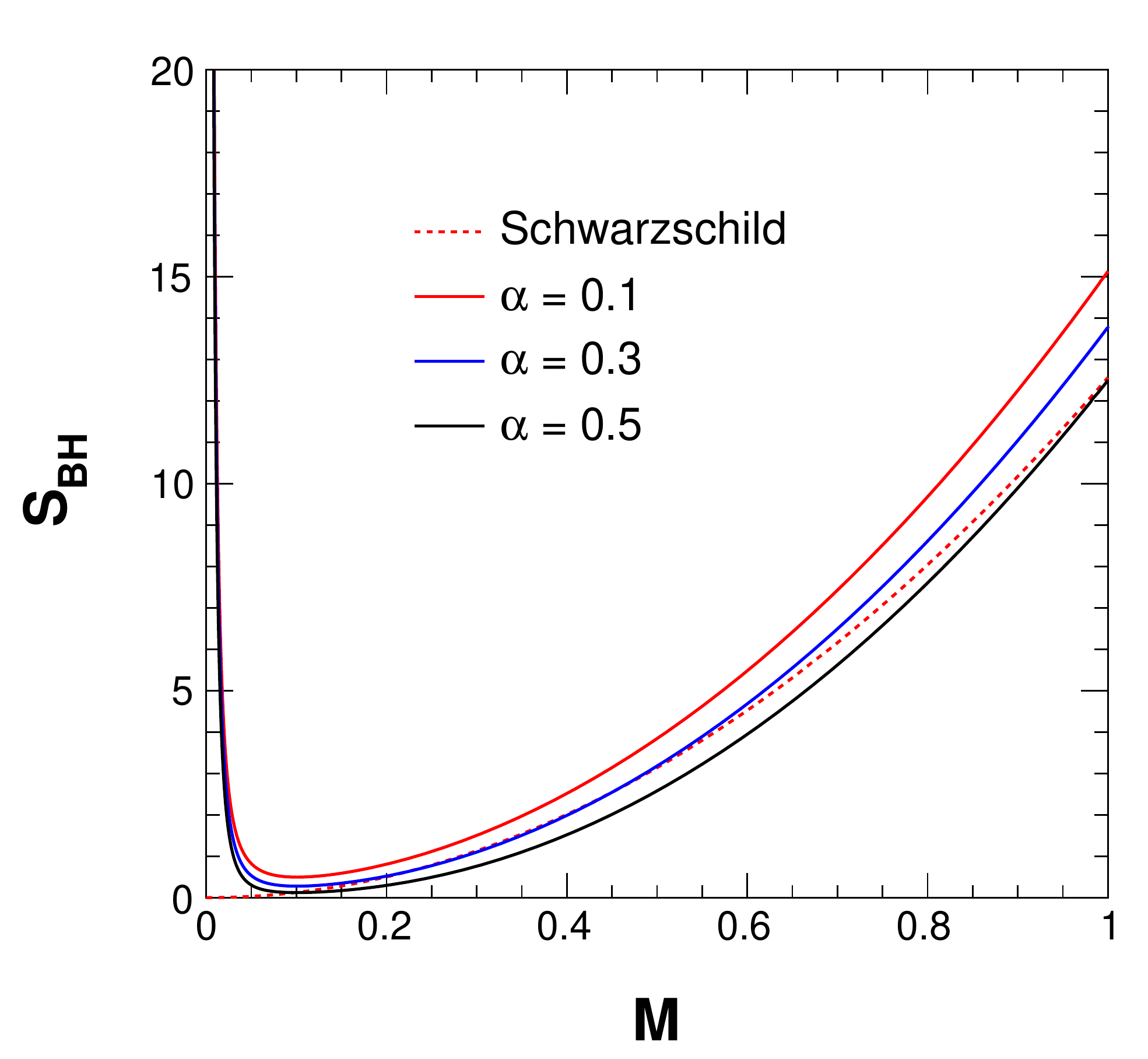}\hspace{0.5cm}
\includegraphics[scale=0.32]{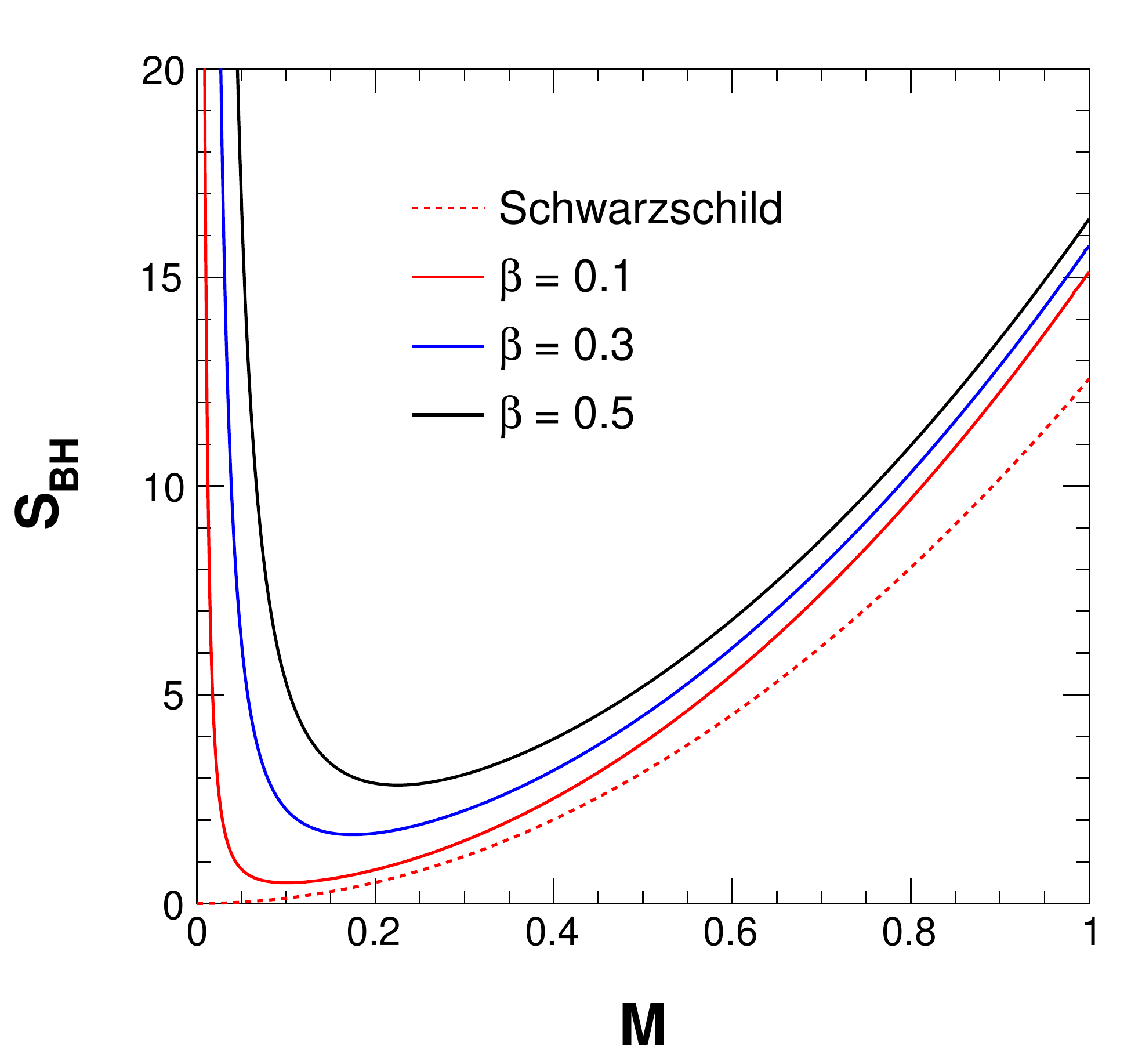}\vspace{0.2cm}
\includegraphics[scale=0.32]{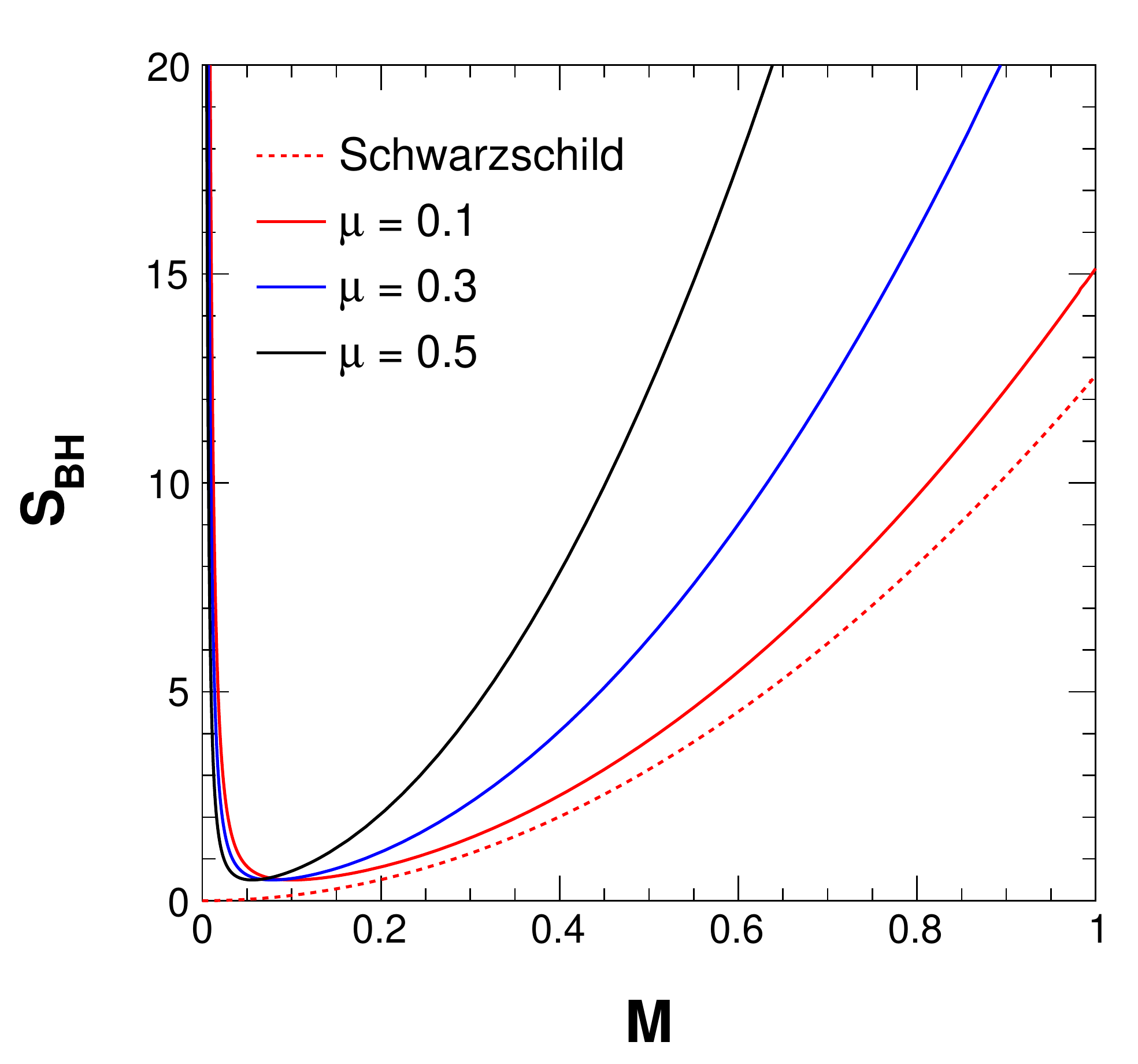}
\caption{ Variation of entropy $S_{BH}$ with mass for various 
parameters of the theory. The first plot is for the variation of entropy 
for different values of $\alpha$, the second one for $\beta$ and the third 
plot is for different values of $\mu$. We use $\beta  = \mu = 0.1$ in the 
first plot, $\alpha = \mu = 0.1$ in the second plot, and 
$\alpha = \beta = 0.1$ in the third plot.}
\label{Fig03}
\end{figure}

Entropy of a black hole is an important parameter that provides an idea about 
the information that falls into the black hole. Entropy is associated with the 
area of the black hole. As more and more matter falls into the black hole, the 
entropy of the black hole keeps on increasing and this is manifested by 
the increase in the horizon area of the black hole. 
 The entropy of a black hole is related to the area of the horizon 
according to the relation,
\begin{equation}
    S_{BH}\equiv\frac{A}{4}=\pi\, r_H^2,
    \label{entropy1}
\end{equation}
where $A$ denotes the area of the black hole. Alternatively, the entropy of a 
black hole can be calculated from the thermodynamic relation \eqref{mtl} as 
follows:
\begin{equation}
  S_{BH}=\int \frac{dM_K}{T_{BH}},
\label{eq:16}
\end{equation}
which gives us the expression for the entropy of the black hole \eqref{Eq:10} as
\begin{equation}
 S_{BH}=\pi r_H^2= \frac{\pi  \Big(\beta  (\mu -1)^2+8 M^2+2 \alpha  (\mu -1) M\Big)^2}{16 (\mu -1)^2 M^2}.
\label{Eq:17}
\end{equation}
 The equivalence of the two expressions \eqref{entropy1} and \eqref{Eq:17} 
again confirms that the first law holds true.
Fig.\ \ref{Fig03} shows the variation of entropy $S_{BH}$ with respect to the
 black hole mass for different values of the parameters of the theory. 
 It is clear from the plots of this figure that entropy of this 
GUP-corrected black hole having topological defects increases with mass for 
all the cases. It is noted that similar to the heat capacity, entropy is also 
independent of $\lambda$. We note that the impacts of increasing $\alpha$ 
and $\beta$ are of reverse nature. For the monopole parameter $\mu$, we see 
that drastic increase of the entropy occurs with a larger value of $\mu$, 
as seen in the third plot. For mass approaching zero, entropy becomes highly 
positive. 

\begin{figure}[h!]
\includegraphics[scale=0.32]{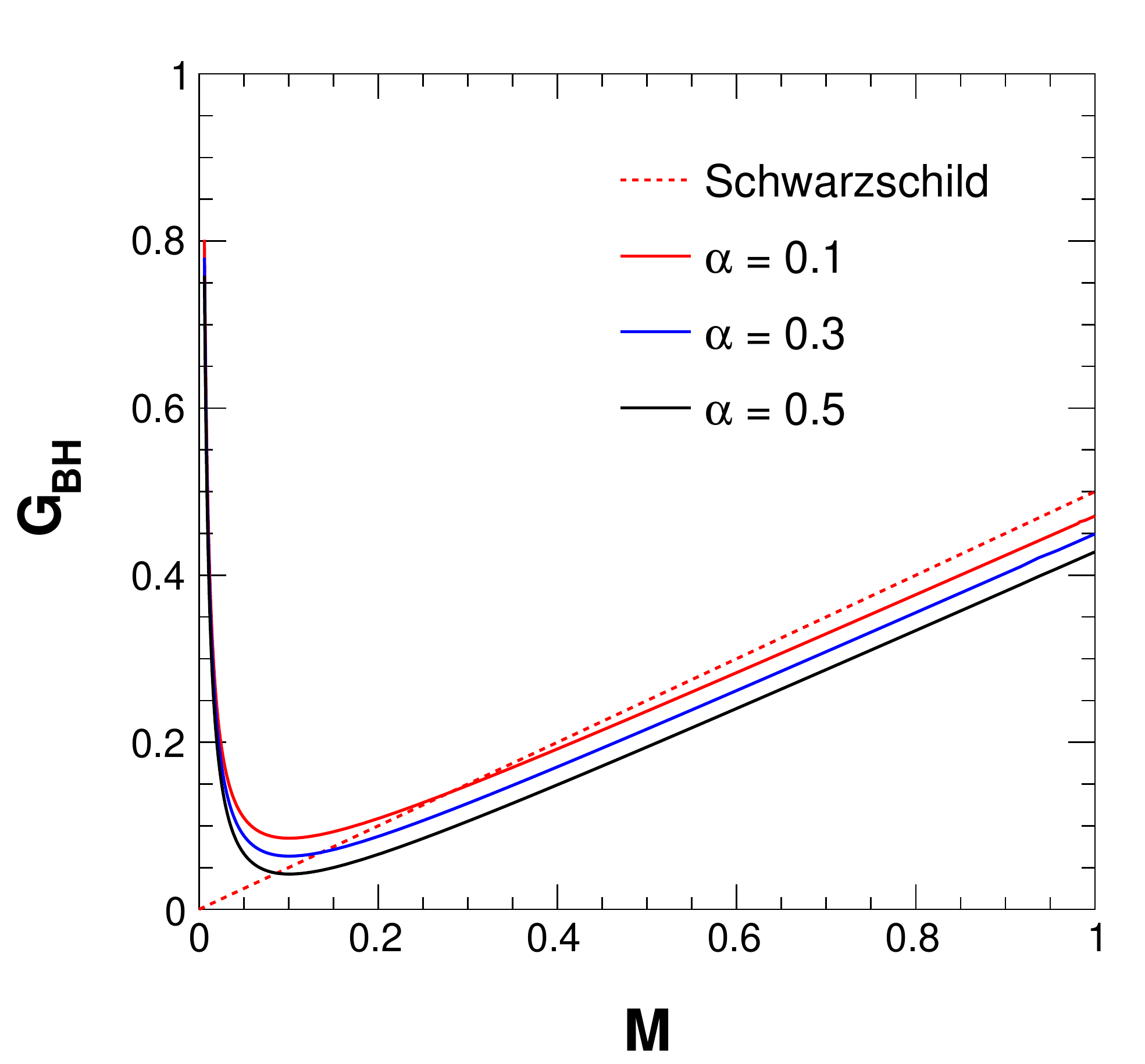}\hspace{0.5cm}
\includegraphics[scale=0.32]{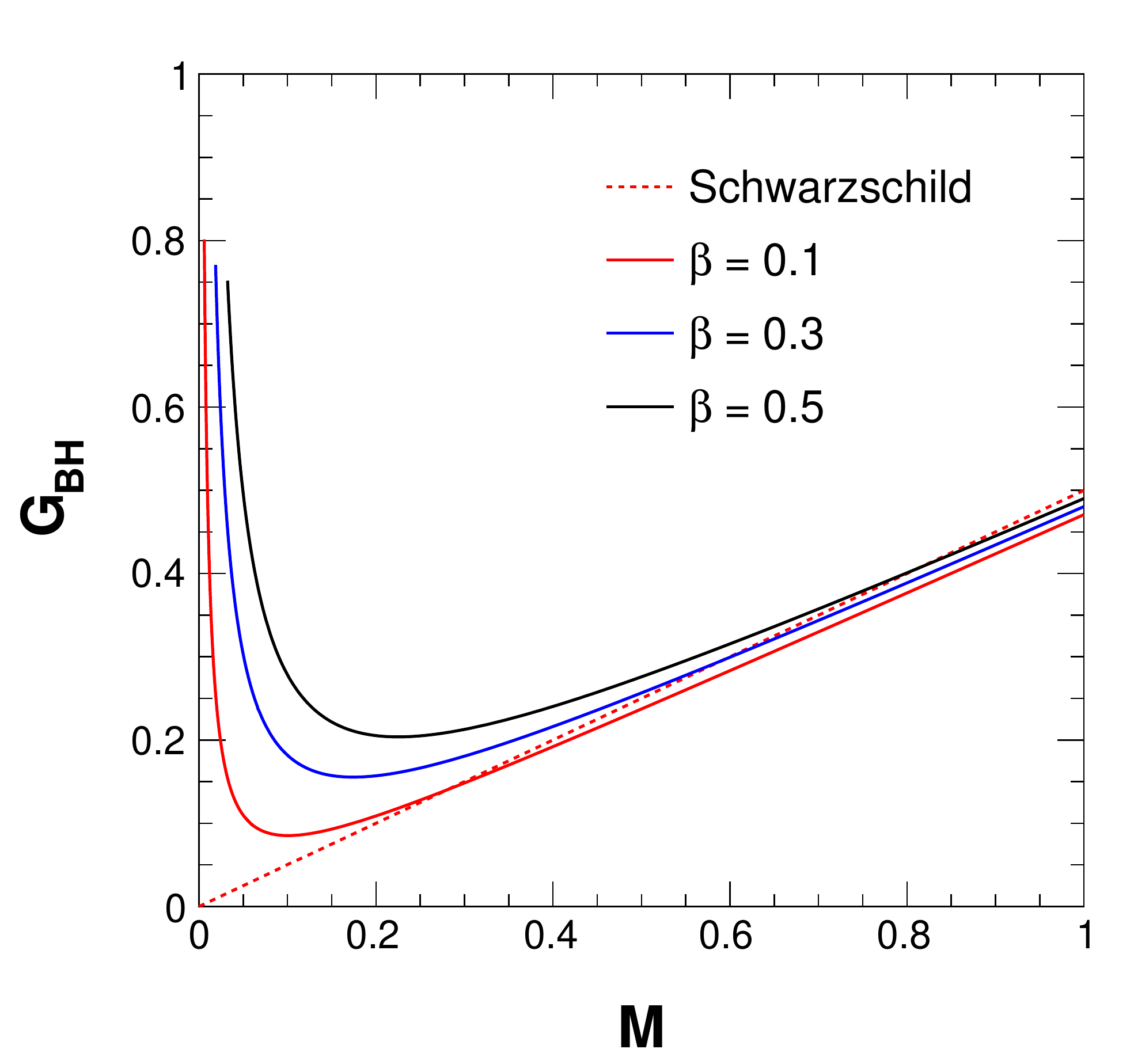}\vspace{0.2cm}
\includegraphics[scale=0.32]{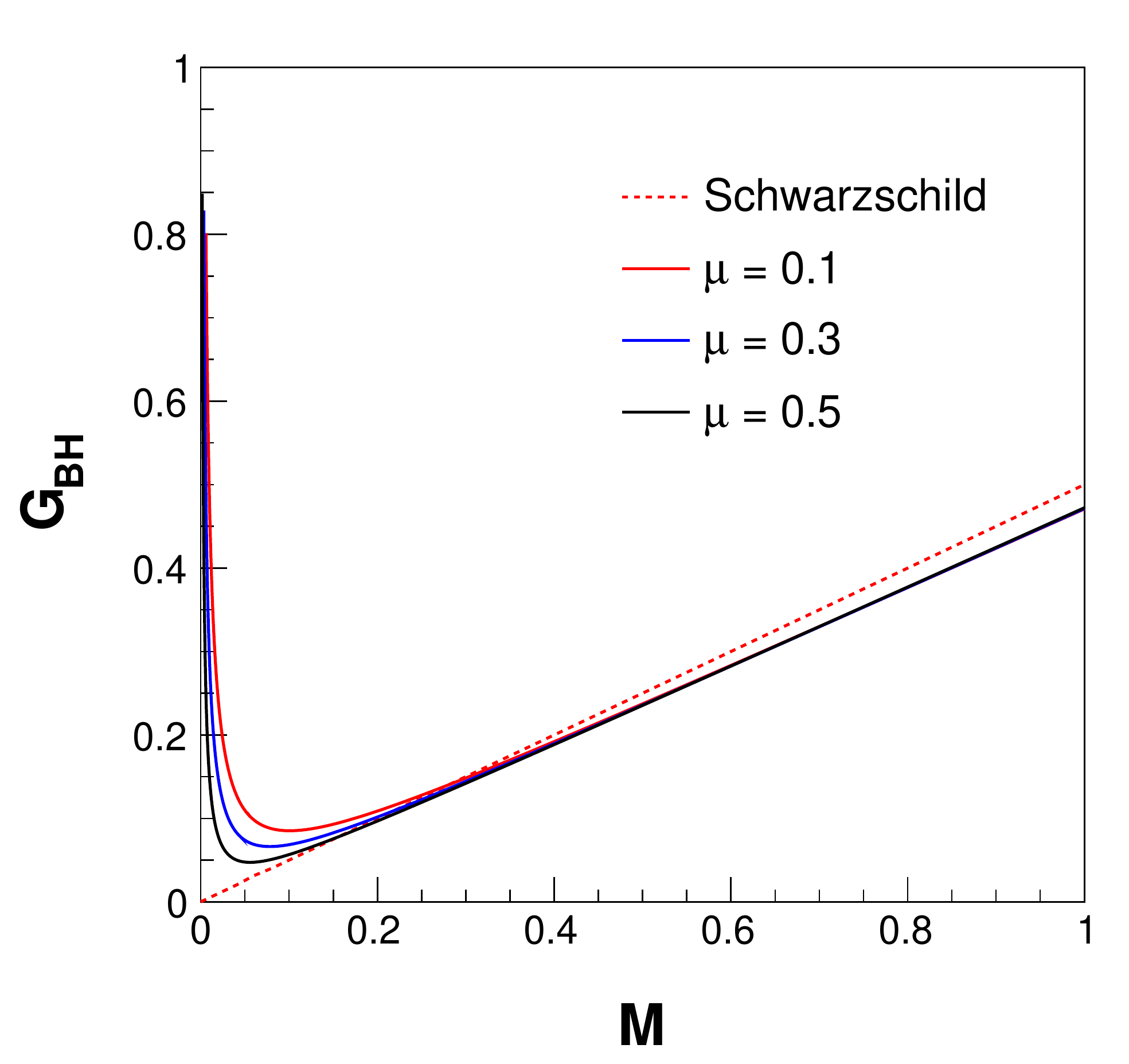}\hspace{0.5cm}
\includegraphics[scale=0.32]{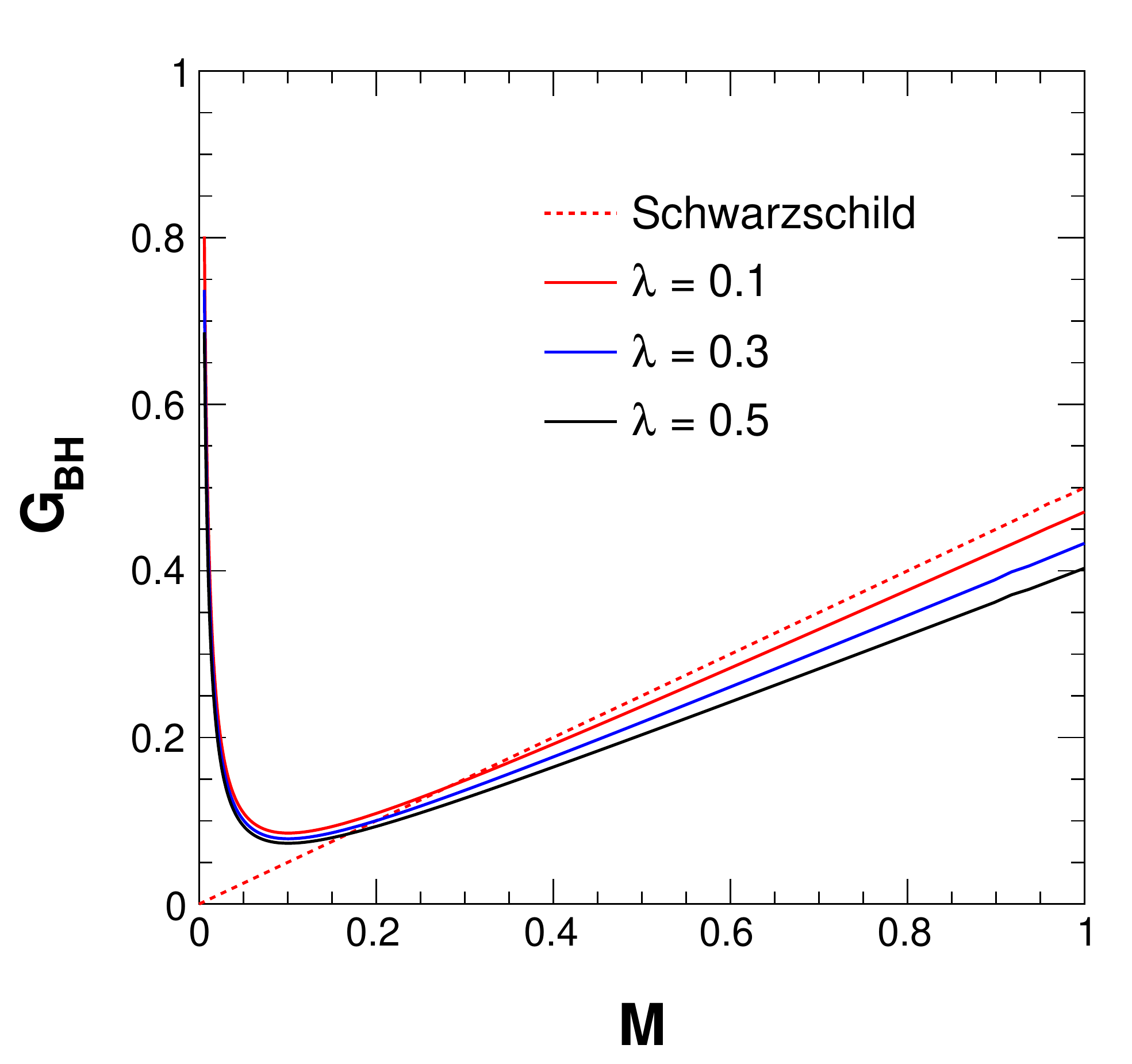}
\caption{ Variation of Gibbs free energy $G_{BH}$ with respect to mass $M$
for various parameters of the theory. The first plot is for the variation of
$G_{BH}$ for different values of $\alpha$, the second one for different $\beta$ values, while the third plot is for different values of $\mu$ and fourth 
plot is for different values of $\lambda$. We use
$\beta = \lambda = \mu = 0.1$ in the first plot, 
$\alpha = \lambda = \mu = 0.1$ in the second plot and 
$\alpha = \beta = \lambda = 0.1$ in the third plot and 
$\alpha = \beta = \mu = 0.1$ in the fourth plot.}
\label{Fig04}
\end{figure}

The Gibb's free energy of a black hole is defined as \cite{50}
\begin{equation}
 G_{BH}=M_K -T_{BH} S_{BH},
\label{Eq:18}
\end{equation}
which gives the expression for the Gibb's free energy for the black hole
of our model as  
\begin{equation}
 G_{BH}=\frac{\beta (\mu -1)^2+8 M^2+2 \alpha (\mu -1) M}{16 \sqrt{\lambda +1} M}.
\label{Eq:19}
\end{equation}
This expression is plotted with respect to mass $M$ for various 
parameters of the model in Fig.\ \ref{Fig04}. As we can see from the 
plots, the variation of Gibbs free energy with mass is similar for $\alpha$, 
$\beta$, $\mu$ as well as $\lambda$. Larger values of parameters $\alpha$ and 
$\lambda$ induce a smaller rise in $G_{BH}$ with $M$. Whereas for $\beta$, this 
trend is opposite and for parameter $\mu$, the various graphs for different 
values of $\mu$ merge together as can be seen.


In the following section, we study the shadow radius of the black 
hole as a function of various parameters of the theory.
Then, in the next section, we derive an upper bound on the values of the 
various parameters of the theory by performing the classical test of 
precession of perihelion of planetary orbits. This kind of analysis was done 
earlier by Casana and his group \cite{17} where they considered 
Schwarzschild-type metric with Lorentz violation parameter involved. Our 
work adds on the global monopole and GUP aspect to the study that provides us 
with interesting insights.

\section{Shadow of the black hole}
\label{sec4}
The shadow formed by a black hole depends on specific parameters of the 
theory and is specified by the photon sphere surrounding the black hole. In 
case of rotating black holes, the shadow is often distorted but 
when we consider spherically symmetric spacetimes, we generally encounter 
spherical shadow radius \cite{1s}. Here our aim is to study the shadow behaviour 
of the black hole in presence of the global monopole, LSB parameter and GUP 
correction. To start off, we form the geodesic equations of a photon moving in 
the black hole spacetime as follows.

The Lagrangian $\mathcal{L}(x,\dot{x})=\frac{1}{2}\,g_{\mu\nu}\dot{x}^{\mu}\dot{x}^{\nu}$ for the case of a spherically symmetric and static spacetime metric 
can be expressed as \cite{shnew02}
\begin{equation}
\mathcal{L}(x,\dot{x})=\frac{1}{2}\left[-f(r)\,\dot{t}^{2}+\frac{1}{g(r)}\,\dot{r}^{2}+r^{2}\left(\dot{\theta}^{2}+\sin^{2}\theta\dot{\phi}^{2}\right)\right],
\end{equation}
where the dot over the variable denotes the derivative with respect to the 
proper time $\tau$ and for our considered black hole spacetime metric 
\eqref{Eq:10}, we have the following form of the metric functions:
 $$f(r)=1-\mu-\frac{2M_{GUP}}{r} \;\;\; \text{and} \;\;\;
g(r) = \frac{f(r)}{(1+\lambda)}.$$ 
We use the Euler-Lagrange equation: 
$\frac{d}{d\tau}\!\left(\frac{\partial\mathcal{L}}{\partial\dot{x}^{\mu}}\right)-\frac{\partial\mathcal{L}}{\partial x^{\mu}}=0$ and choose the equatorial 
plane, i.e. $\theta=\pi/2$, in order to derive the conserved quantities of the
system, viz. energy and angular momentum. Two killing vectors 
$\partial/\partial \tau$ and $\partial/\partial \phi$ yield conserved energy 
$\mathcal{E}$ and angular momentum $L$ for the considered case as
\begin{equation}
\mathcal{E}=f(r)\,\dot{t},\quad L=r^{2}\dot{\phi}.
\end{equation}
In case of photon, the geodesic equation leads to the relation,
\begin{equation}\label{eq22}
-f(r)\,\dot{t}^{2}+\frac{(1+\lambda)}{f(r)}\,\dot{r}^{2}+r^{2}\dot{\phi}^{2} = 0.
\end{equation}
We utilise the conserved quantities i.e.\ $\mathcal{E}$ and $L$ in the above 
Eq.\ \eqref{eq22}, which is required for obtaining the photon's orbital 
equation as given by \cite{shnew01}
\begin{equation}\label{eff}
\left(\frac{dr}{d\phi}\right)^{2}=\frac{r^{4}}{(1+\lambda)}\left[\frac{\mathcal{E}^{2}}{L^{2}}-\frac{f(r)}{r^2}\right].
\end{equation}
Defining the right hand side of the above equation as an effective potential
$V_{eff}$, i.e.
\begin{equation}
V_{eff}=\frac{r^{4}}{(1+\lambda)}\left[\frac{\mathcal{E}^{2}}{L^{2}}-\frac{f(r)}{r^{2}}\right]
\end{equation}
the equation can be expressed in a compact form as 
\begin{equation}
\left(\frac{dr}{d\phi}\right)^{2}=V_{eff}.
\end{equation}
Moreover, Eq.\ \eqref{eff} can be rewritten in the form of a radial equation
as given by
\begin{equation}
\dot{r}^2 + V_r(r) = \frac{\mathcal{E}^2}{1+\lambda},
\end{equation}
where $V_r(r)$ is a new potential, we refer it as the reduced potential, which 
has the following form:
\begin{equation}
V_r(r) = \frac{f(r)L^2}{ r^2 (1+\lambda)}.
\end{equation}
It is to be noted that since the angular momentum $L$ is a conserved quantity, 
it remains constant throughout and hence it will not have any impact on the 
overall behaviour of the reduced potential $V_r(r)$. Thus for the 
simplicity we will consider $L=1$ in our calculation of this potential. 
Further, as this potential governs the radial motion of photons in the black 
hole spacetime, the study of the behaviour of this potential with respect to 
the radial distance $r$ would be the realistic approach to understand the 
nature of the photon sphere around the black hole spacetime we have 
considered. 
\begin{figure}[h!]
\vspace{0.2cm}
\centering{
\includegraphics[scale=0.35]{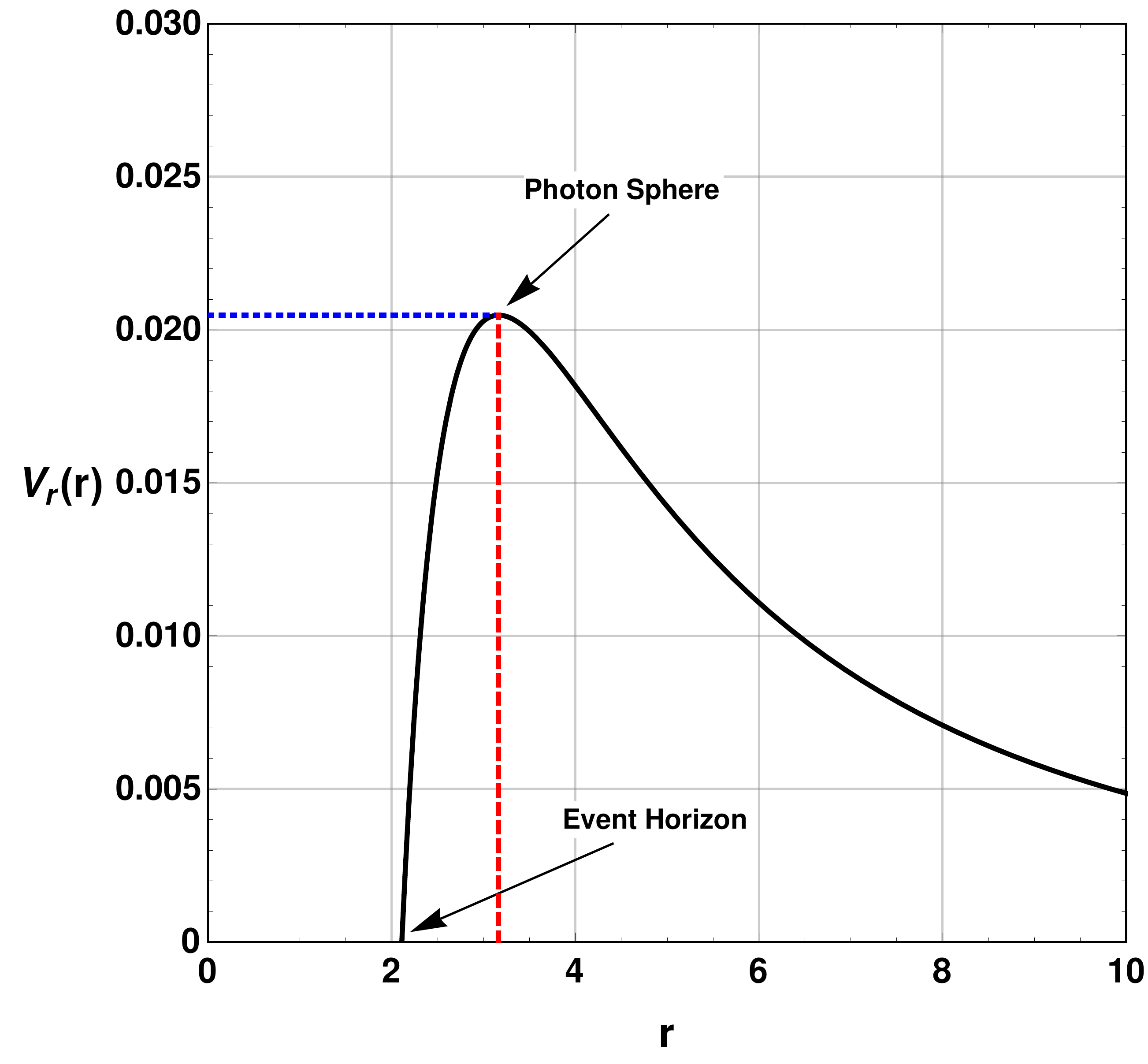} 
}
\caption{Variation of the reduced potential $V_r(r)$ with respect to distance 
$r$ as obtained by using $M=1$, $\alpha=0.8$, $\beta = 0.05$, $\lambda = 0.3$ and 
$\mu=0.2$.}
\label{Shadow00a}
\end{figure}

\begin{figure}[h!]
\centering{
\includegraphics[scale=0.55]{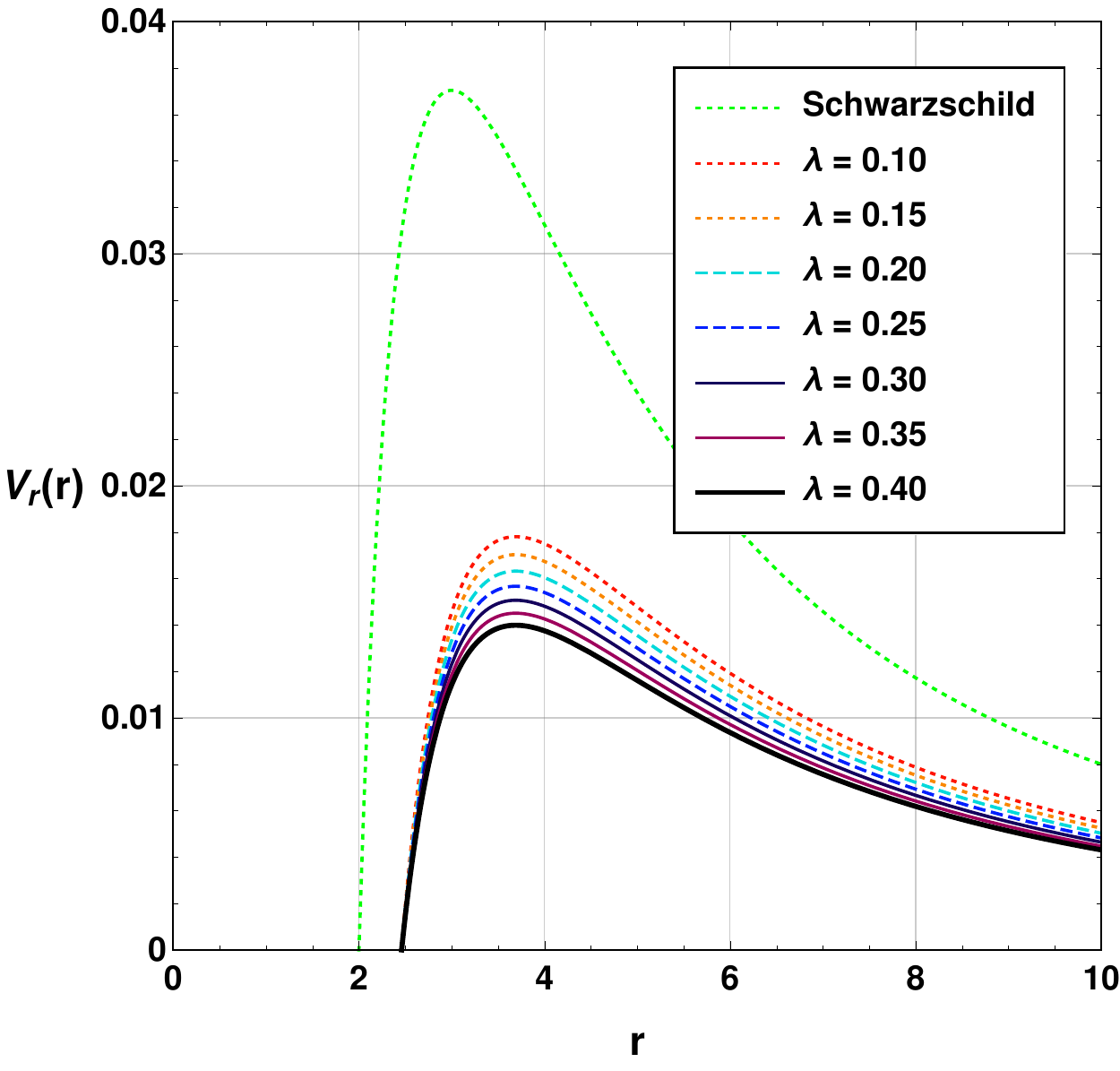} \hspace{0.5cm}
\includegraphics[scale=0.55]{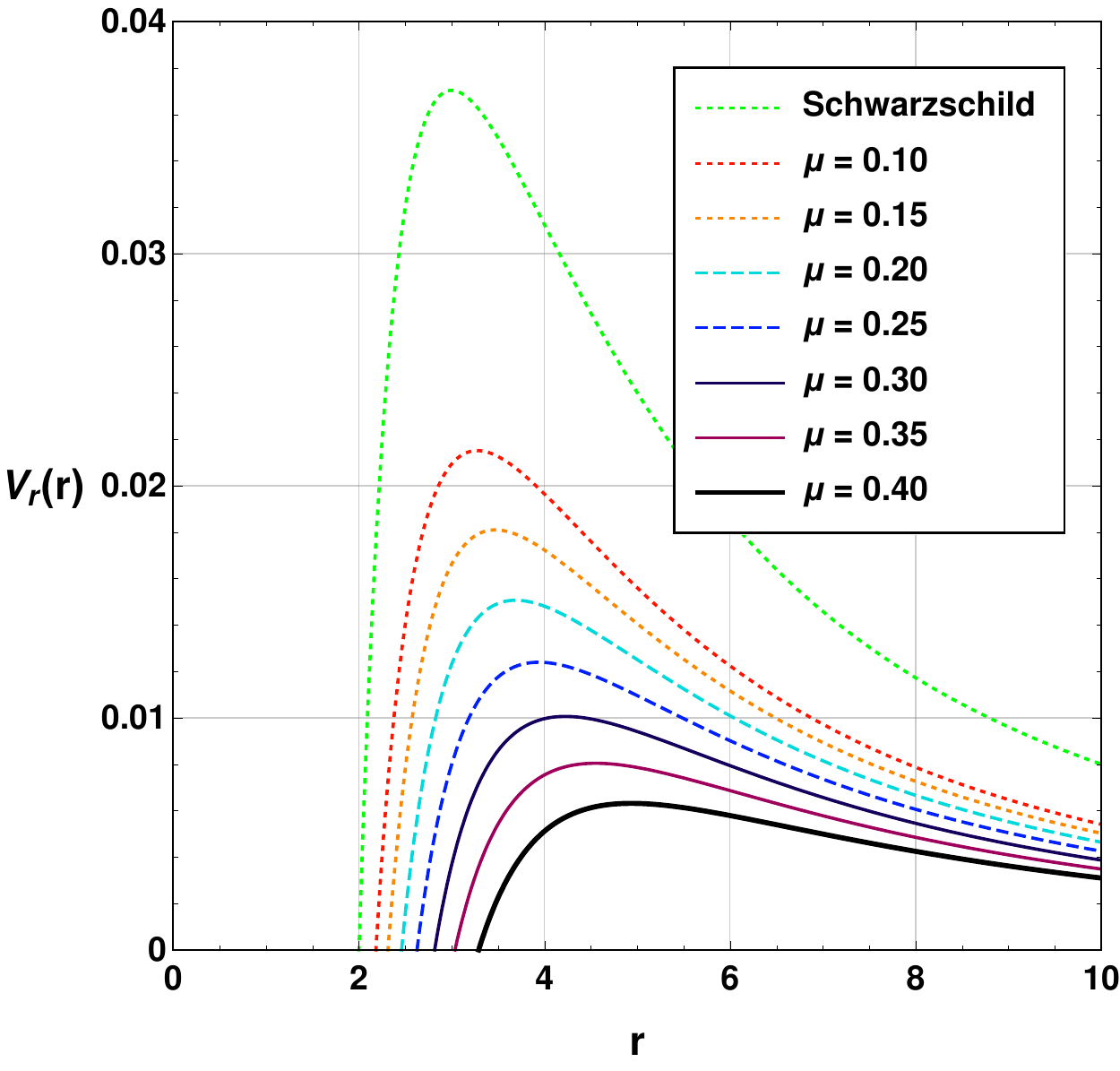}
}
\caption{Variation of the reduced potential $V_r(r)$ with respect to distance 
$r$ for different values of LSB parameter $\lambda$ (left plot) and global
monopole term $\mu$ (right plot). We use $M=1$, $\alpha=0.1$, 
$\beta = 0.05$ and $\mu=0.2$ in the left plot, and $M=1$, $\alpha=0.1$, 
$\beta = 0.05$ and $\lambda = 0.3$ in the right plot.}
\label{Shadow00b}
\end{figure}

\begin{figure}[h!]
\centering{
\includegraphics[scale=0.55]{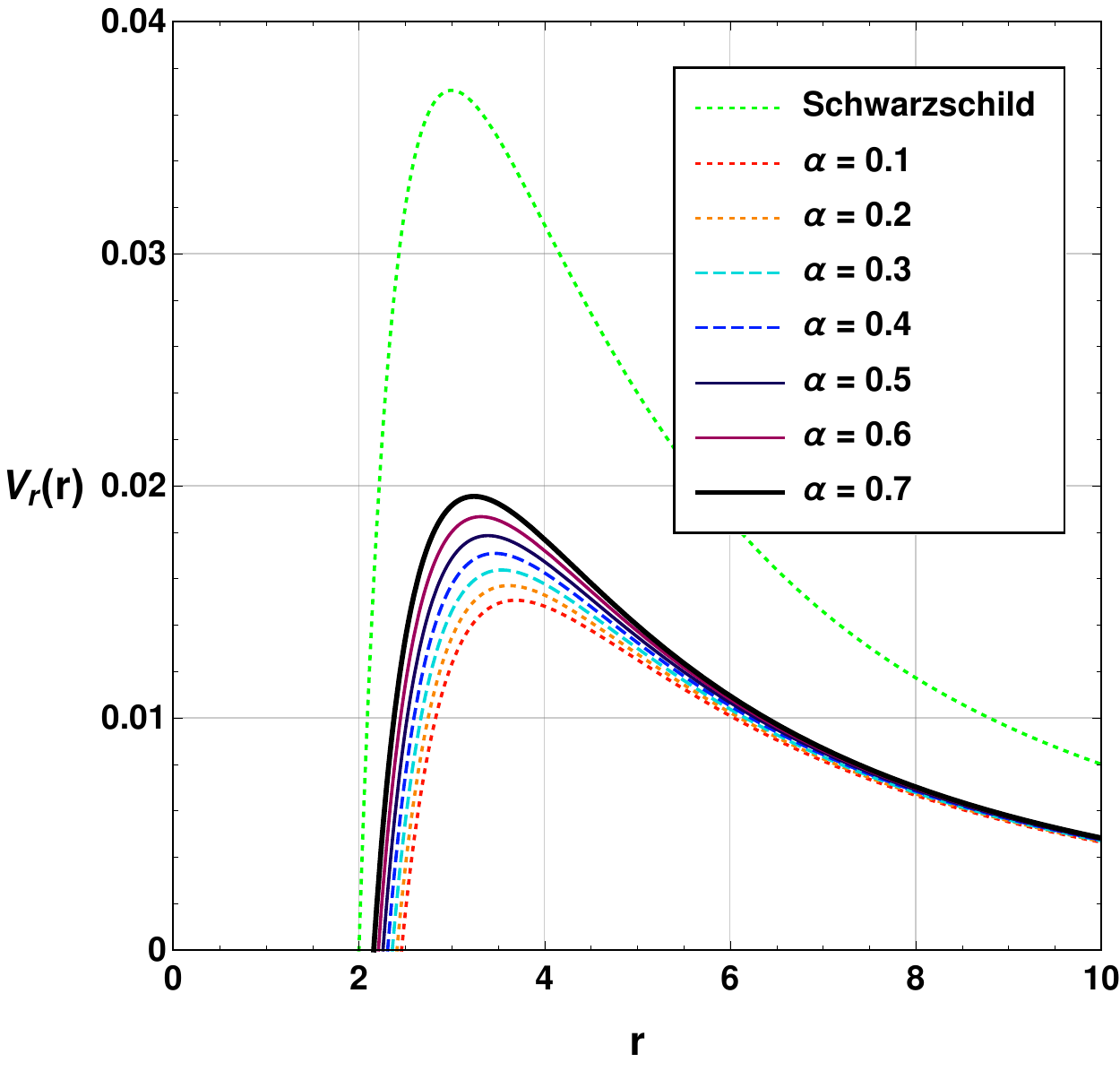} \hspace{0.5cm}
\includegraphics[scale=0.55]{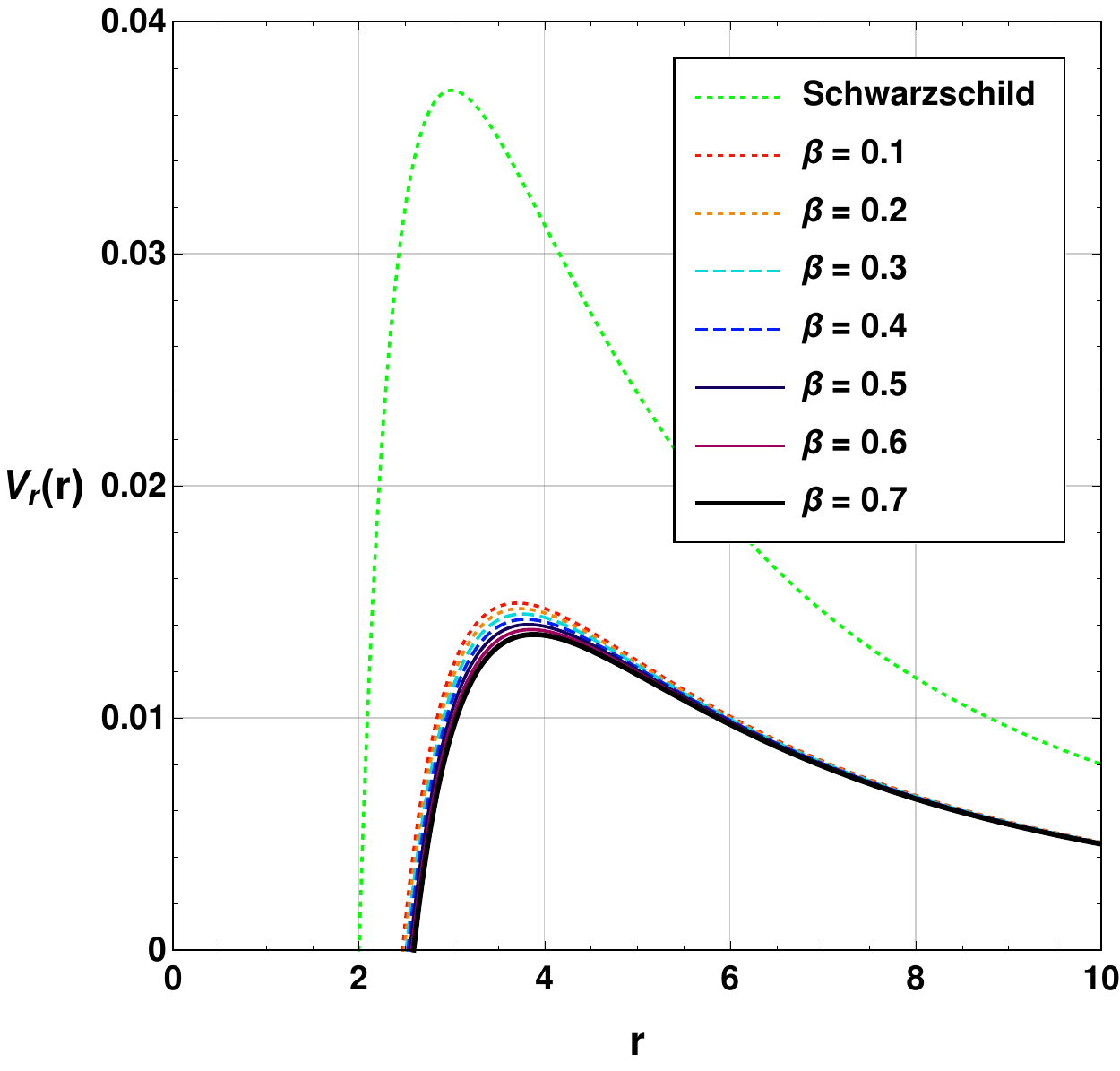}
}
\caption{Variation of the reduced potential $V_r(r)$ with respect to distance 
$r$ for different values GUP parameters $\alpha$ (left plot) and $\beta$ 
(right plot). We use $M=1$, $\beta = 0.05$, $\lambda = 0.3$ and $\mu=0.2$ in 
the left plot, and $M=1$, $\alpha=0.1$, $\mu = 0.2$ and $\lambda = 0.3$ in the right
plot.}
\label{Shadow00c}
\end{figure}

In Fig.\ \ref{Shadow00a} we show the behaviour of this reduced potential 
$V_r(r)$ with respect to distance $r$. In the figure the peak position of the 
potential curve represents the photon sphere, whose radius is measured as a
parallel distance from the $y$-axis to peak position and is distinctly shown 
in the plot. On the left panel of Fig.\ \ref{Shadow00b} we show the 
behaviour of this potential with respect to $r$ for different values of LSB 
parameter $\lambda$. One can see that with an increase in the LSB parameter 
$\lambda$, the peak value of the potential decreases significantly. However, 
this parameter does not have any impact on the photon sphere size of the black 
hole as the peak position of the potential does not change its position with
respect to $r$. On the right panel of Fig.\ \ref{Shadow00b} we plot 
the potential with respect to $r$ for different values of the global monopole 
term $\mu$. It shows that the global monopole parameter can have a significant 
impact on the potential as well as the photon sphere of the black hole. With 
an increase in the value of $\mu$, the peak value of the potential decreases 
drastically while the photon sphere size increases noticeably. In 
Fig.\ \ref{Shadow00c} we show the behaviour of the potential for 
different values of the GUP parameters. On the left panel we consider 
different values of the first GUP parameter $\alpha$, where we have seen that 
with an increase in value of $\alpha$, the peak value of the potential 
increases, unlike the scenario for the LSB parameter $\lambda$ and monopole 
parameter $\mu$. The photon sphere size decreases with an increase in the 
value of $\alpha$. On the right panel of Fig.\ \ref{Shadow00c}, we show 
the impacts of the second GUP parameter $\beta$ on the potential of the black 
hole. Here, we see that with an increase in the value of the parameter 
$\beta$, the photon sphere increases in size while the peak value of the 
potential decreases. 
From this graphical analysis we have seen that the size of the photon sphere 
basically depends on three parameters only, viz., monopole term $\mu$, GUP 
parameters $\alpha$ and $\beta$. The second GUP parameter $\beta$ has the 
smallest impact on the potential of the black hole, as seen from the analysis.

Now, we move to the analysis of the shadow of the black hole. To obtain the
shadow of the black hole, we consider the trajectory's turning point, denoted 
by $r=r_{ph}$, which is in fact the radius of the photon sphere or light ring
around the black hole. 
At this turning point, the conditions that must be satisfied are 
\cite{18, synge, Luminet:1979nyg}
\begin{equation} 
\left.\frac{dr}{d\phi}\right|_{r_{ph}}\!\!\!\!\!\!\!=0\;\; \text{or}\;\; \left.V_{eff}\right|_{r_{ph}}\!\!\!\!=0,\;\; \text{and}\;\;\; \left.\frac{d^2r}{d\phi^2}\right|_{r_{ph}}\!\!\!\!\!\!\!=0
\;\; \text{or}\;\; \left.V_{eff}^{\prime}\right|_{r_{ph}}\!\!\!\!=0.
\end{equation} 
The impact parameter $b$ at the turning point obtained from the first 
condition is
\begin{equation}
\frac{1}{b_{crit}^{2}}=\frac{f(r_{ph})}{r_{ph}^{2}}.
\label{impact}
\end{equation}
Here, the parameter $b$ is defined as $b=L/\mathcal{E}$. From the second one of 
above conditions one can find the radius of the photon sphere $r_{ph}$ by 
solving the equation: 
\begin{equation}
\left.\frac{d}{dr}\,\mathcal{A}(r)\right|_{r_{ph}}\!\!\!\!\!\!\! = 0,
\end{equation}
which can be written explicitly as
\begin{equation}
\frac{f^{\prime}(r_{ph})}{f(r_{ph})}-\frac{h^{\prime}(r_{ph})}{h(r_{ph})}=0,
\label{photon}
\end{equation}
where $\mathcal{A}(r)=h(r)/f(r)$ with $h(r)=r^{2}$. The analysis of 
Eqs.\ \eqref{impact} and \eqref{photon} shows that the radius of the photon 
sphere is at $r_{ph}=3M_{GUP}/(1-\mu)$ and the critical impact parameter is 
$b_{crit}=3\sqrt{3}M_{GUP}/\sqrt{1-\mu^3+3\mu^2-3\mu}$. In the absence of the 
GUP corrections and global monopole, the radius of the photon sphere is 
$r_{ph}=3M$ and the critical impact parameter is $b_{crit}=3\sqrt{3}M$ which 
corresponds to a Schwarzschild black hole.

In terms of the function $\mathcal{A}(r)$, Eq.\ \eqref{eff} can be rewritten 
with Eq.\ \eqref{impact} as 
\begin{equation}
\left(\frac{dr}{d\phi}\right)^{\!2}=\frac{h(r)f(r)}{1+\lambda}\left(\frac{\mathcal{A}(r)}{\mathcal{A}(r_{ph})}-1\right).
\label{eq33}
\end{equation}
This equation can be used to calculate the shadow radius. For this purpose, if 
we consider that $\alpha$ is the angle between the light rays from a static 
observer at $r_0$ and the radial direction of the photon sphere, then the
angle $\alpha$ can be found as \cite{Perlick:2021aok, 18}
\begin{equation}
\cot\alpha=\frac{\sqrt{(1+\lambda)}}{\sqrt{f(r)h(r)}}\left.\frac{dr}{d\phi}\right|_{r\,=\,r_{0}}\!\!\!\!\!\!\!\!\!\!\!.
\end{equation}
With Eq.\ \eqref{eq33}, above equation can be expressed as 
\begin{equation}
\cot^{2}\!\alpha=\frac{\mathcal{A}(r_{0})}{\mathcal{A}(r_{ph})}-1.
\end{equation}
Using the relation $\sin^{2}\!\alpha=1/(1+\cot^{2}\!\alpha)$, above equation 
can be rewritten as
\begin{equation}
\sin^{2}\!\alpha=\frac{\mathcal{A}(r_{ph})}{\mathcal{A}(r_{0})}.
\end{equation}
Substituting the actual form of $\mathcal{A}(r_{ph})$ from Eq.\ \eqref{impact} 
and $\mathcal{A}(r_{0}) = r_0^2/f(r_0)$,
the shadow radius of the black hole for a static observer at $r_{0}$
is found as \cite{15s} 
\begin{equation}
R_{s}=r_{0}\sin\alpha=\sqrt{\frac{r_{ph}^2f(r_{0})}{f\left(r_{ph}\right)}}.
\end{equation}
Again, for a static observer at large distance, 
i.e.\ at $r_0 \rightarrow \infty$, $f(r_0) \rightarrow 1$, so for such an 
observer the shadow radius $R_s$ becomes, 
\begin{equation}
R_{s} = \frac{r_{ph}}{\sqrt{f(r_{ph})}}.
\end{equation}
Finally, the apparent shape of the shadow can be found by the stereographic 
projection of the shadow from the black hole's plane to the observer's image
plane with coordinates $(X,Y)$. These coordinates are defined as \cite{shnew03} 
\begin{align}
 X & =\lim_{r_{0}\rightarrow\infty}\left(-\,r_{0}^{2}\sin\theta_{0}\left.\frac{d\phi}{dr}\right|_{r_{0}}\right),\\[5pt]
Y & =\lim_{r_{0}\rightarrow\infty}\left(r_{0}^{2}\left.\frac{d\theta}{dr}\right|_{(r_{0},\theta_{0})}\right),
\end{align}
where $\theta_{0}$ is angular position of the observer with respect to the 
black hole's plane. 
\begin{figure}[h!]
\vspace{0.5cm}
\centering{
\includegraphics[scale=0.5]{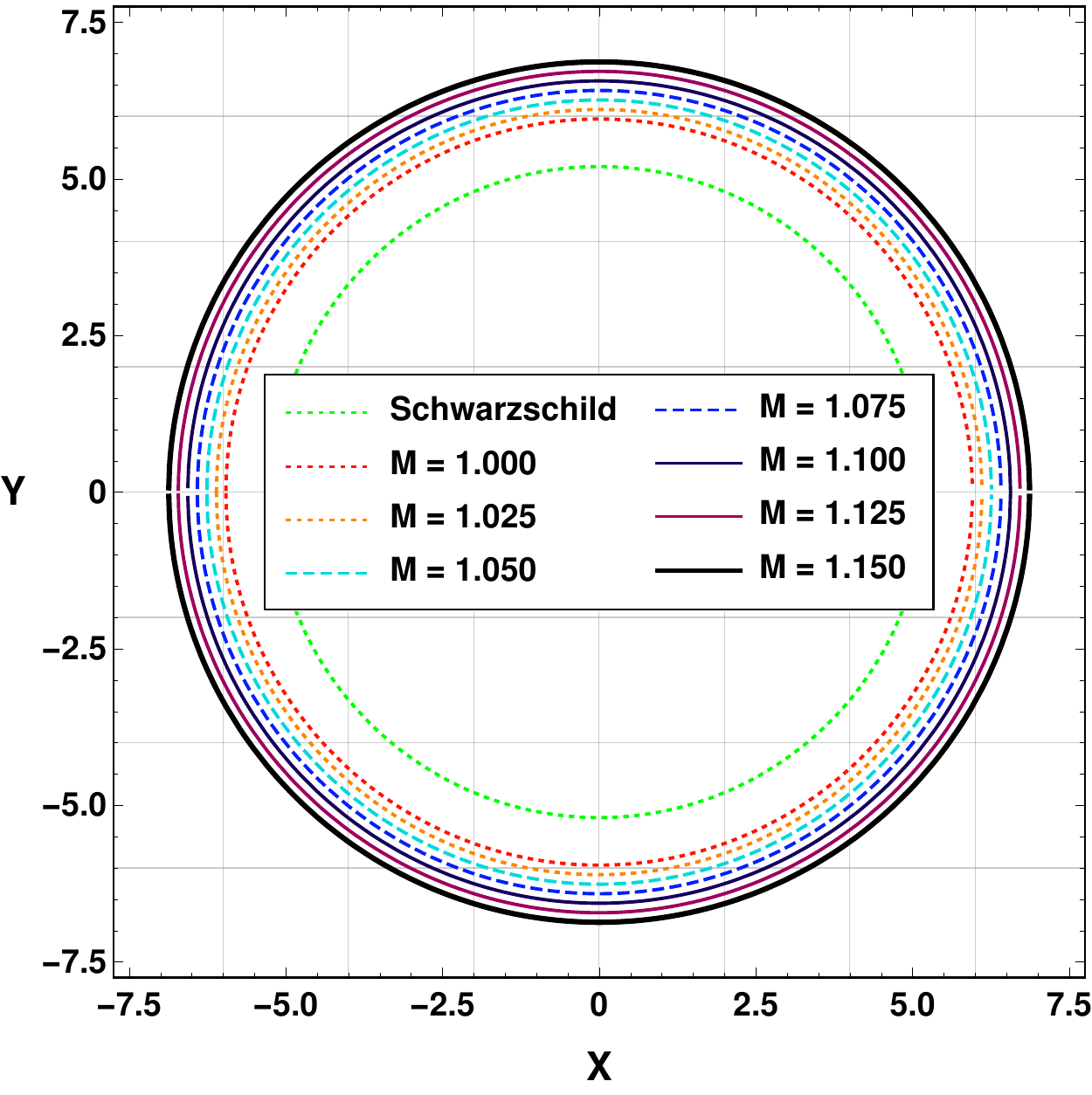} \hspace{1cm}
\includegraphics[scale=0.5]{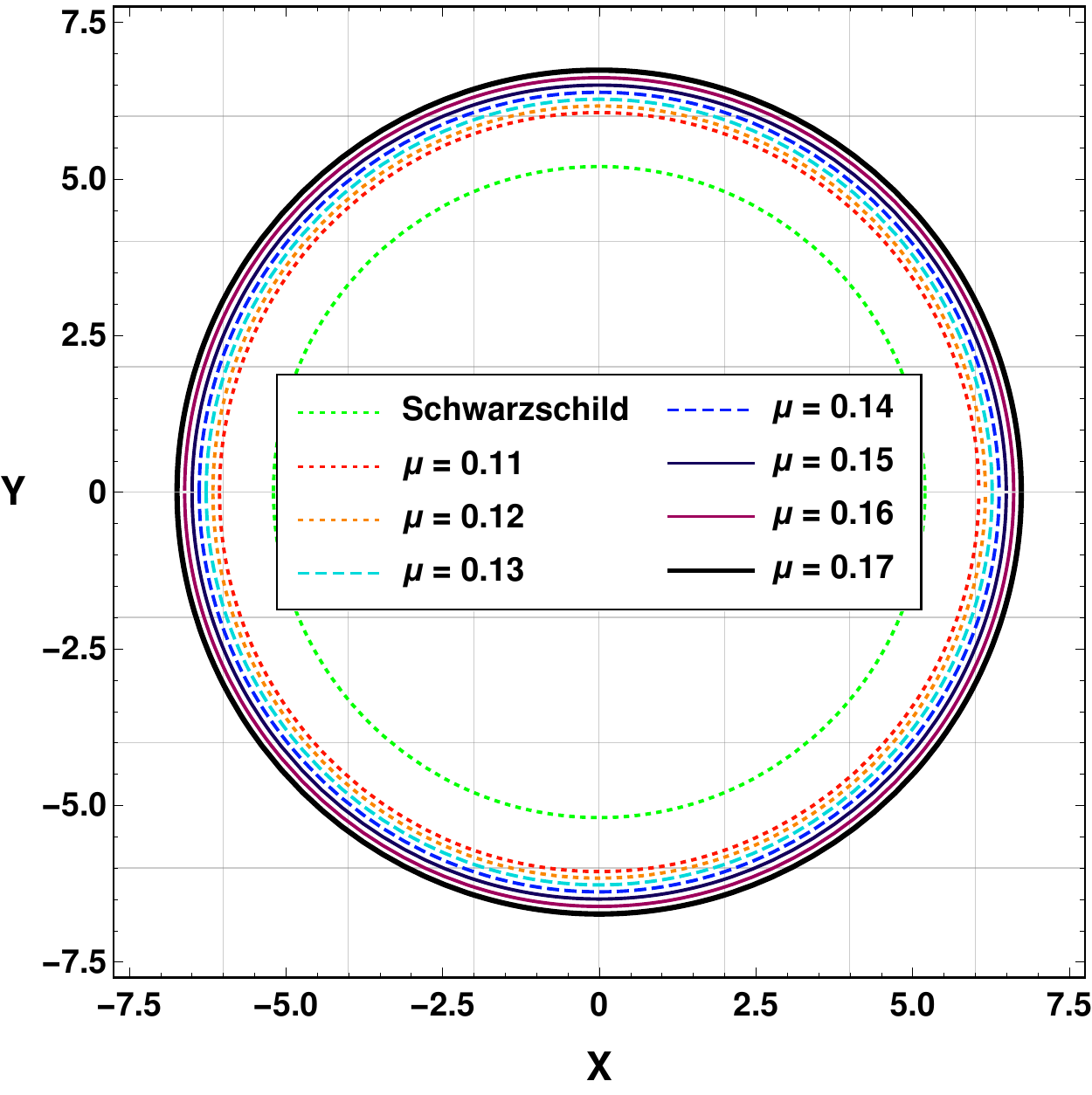}
}
\caption{Stereographic projection of the black hole shadow in the observer 
image plane obtained by using (a) $\alpha = 0.1$, $\beta = 0.01$ and 
$\mu=0.1$ on the left panel and (b) $\alpha = 0.1$, $\beta = 0.01$ and 
$M=1$ on the right panel.}
\label{Shadow01}
\end{figure}

\begin{figure}[h!]
\centering{
\includegraphics[scale=0.5]{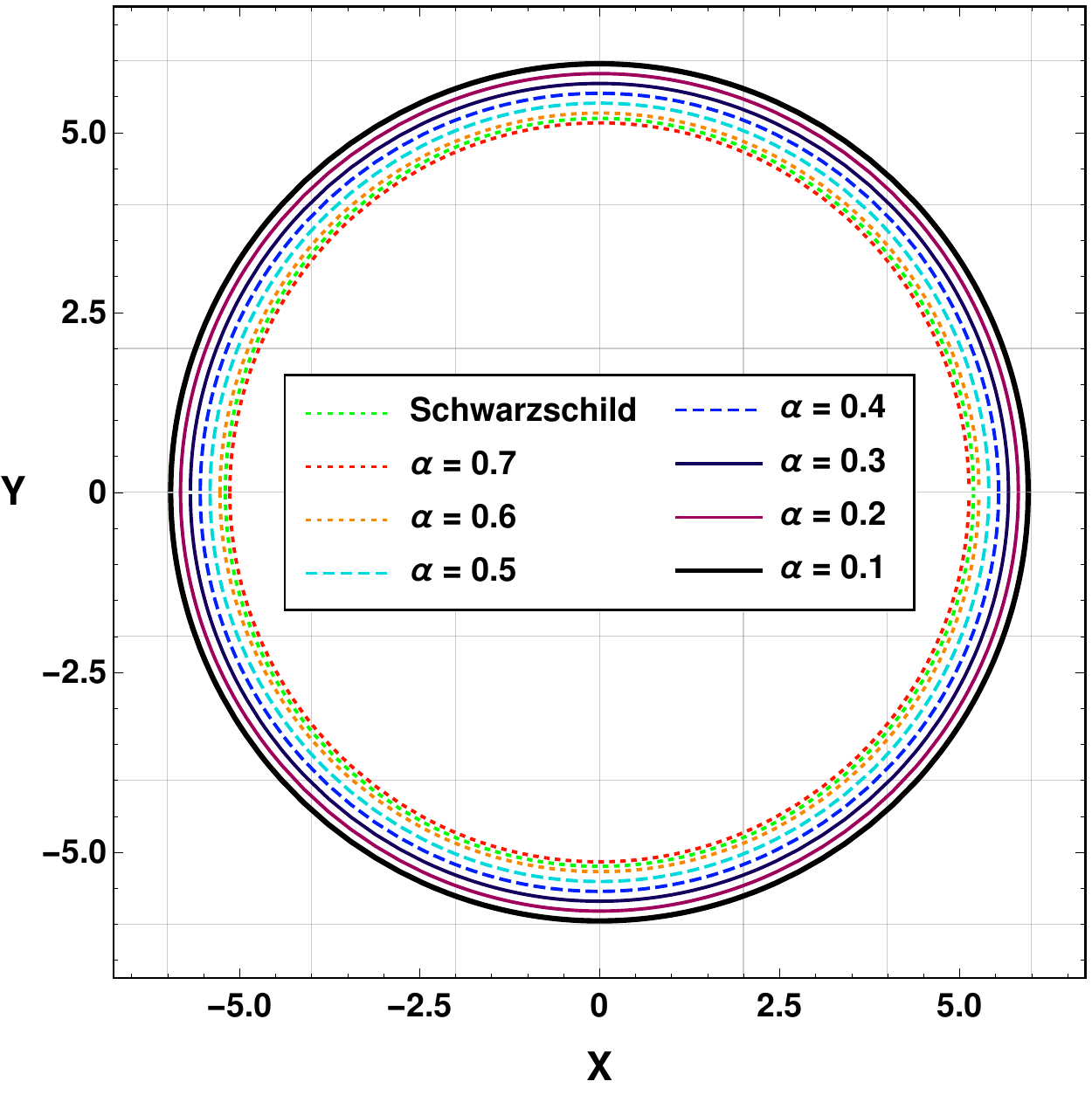} \hspace{1cm}
\includegraphics[scale=0.5]{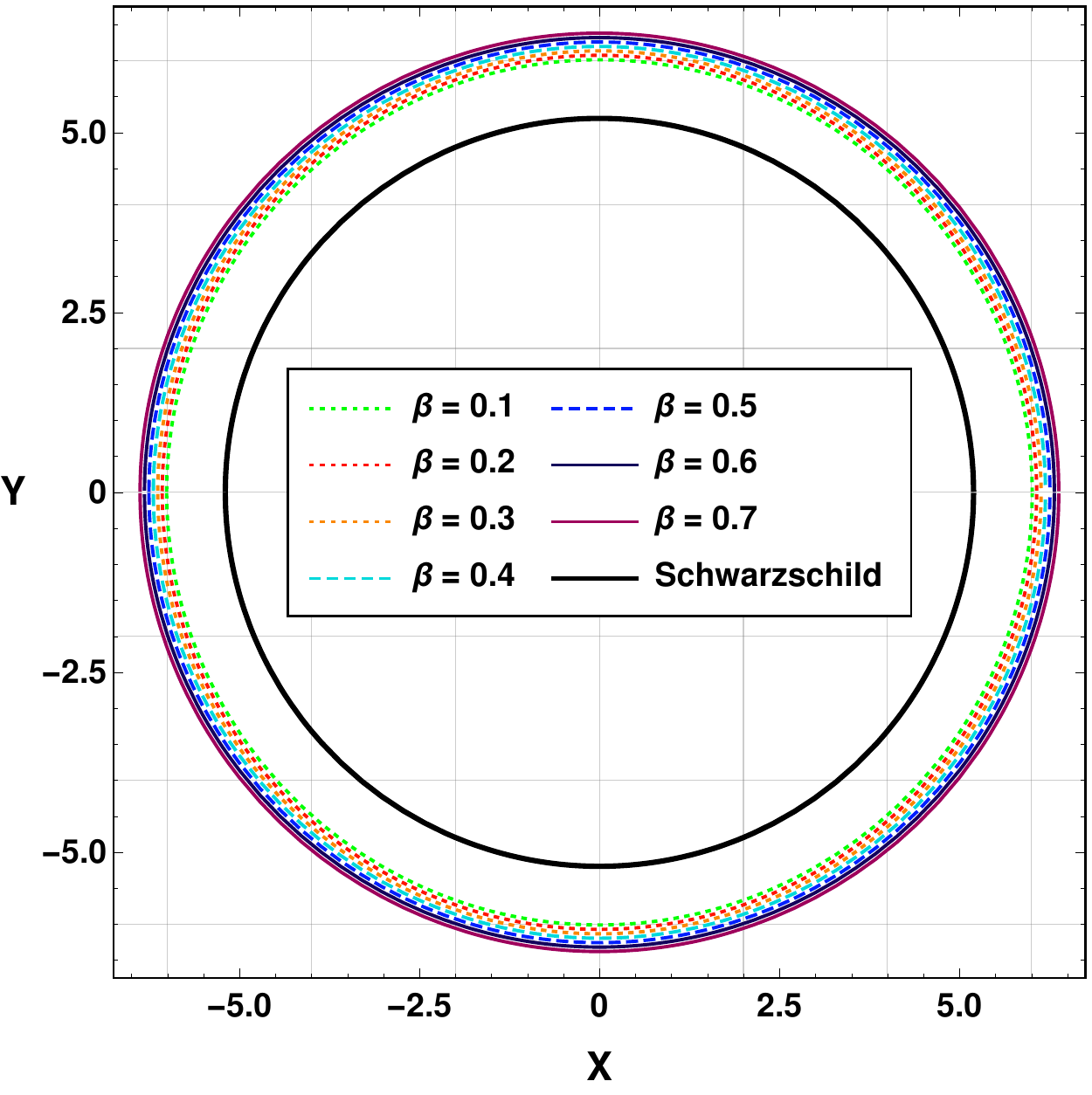}
}
\caption{Stereographic projection of the black hole shadow in the observer 
image plane obtained by using (a) $M=1$, $\beta = 0.01$ and $\mu=0.1$ on the 
left panel and (b) $M=1$, $\alpha = 0.1$ and $\mu = 0.1$ on the right panel.}
\label{Shadow02}
\end{figure}

\begin{figure}[h!]
\centering{
\includegraphics[scale=0.5]{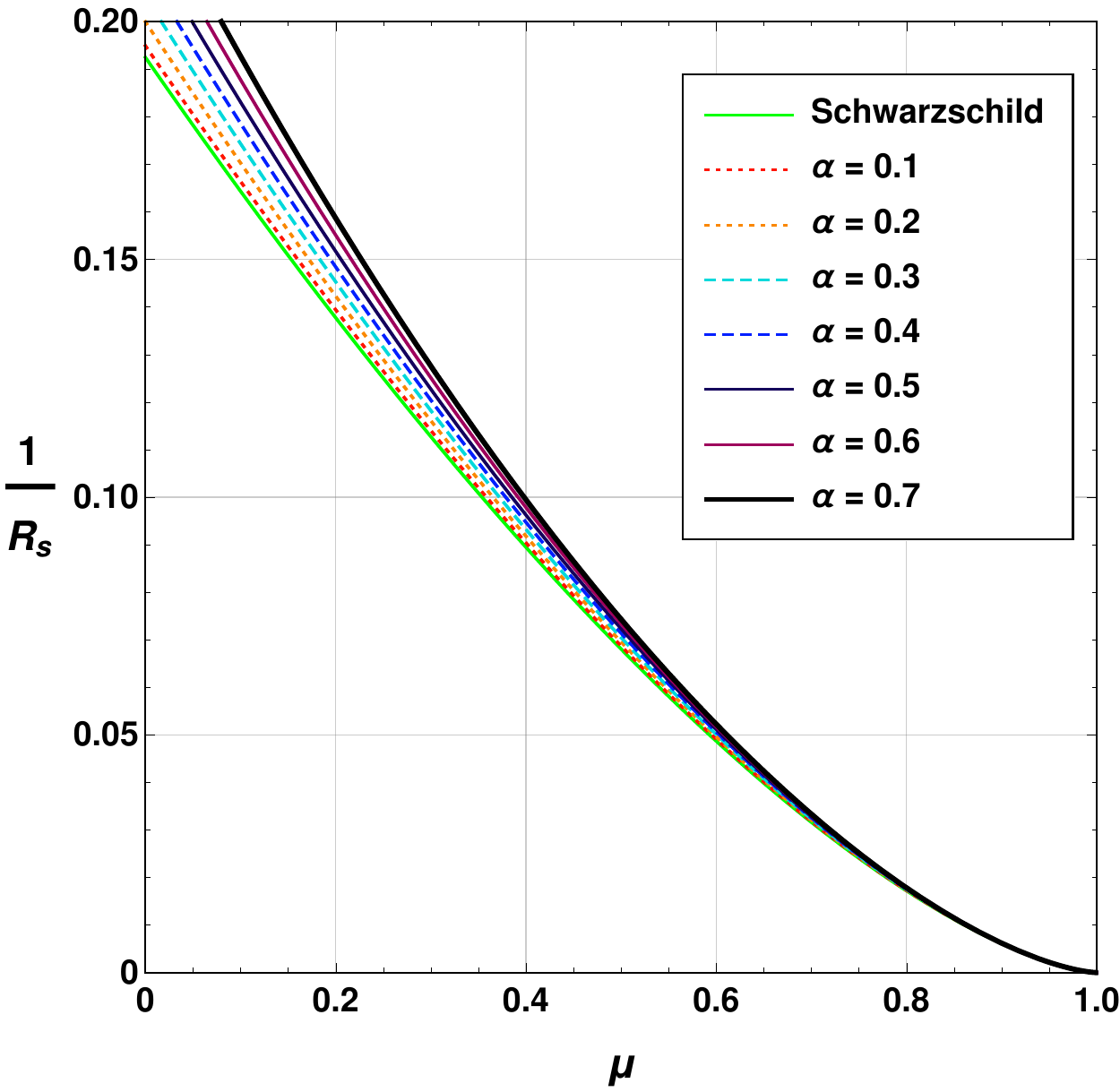} \hspace{1cm}
\includegraphics[scale=0.5]{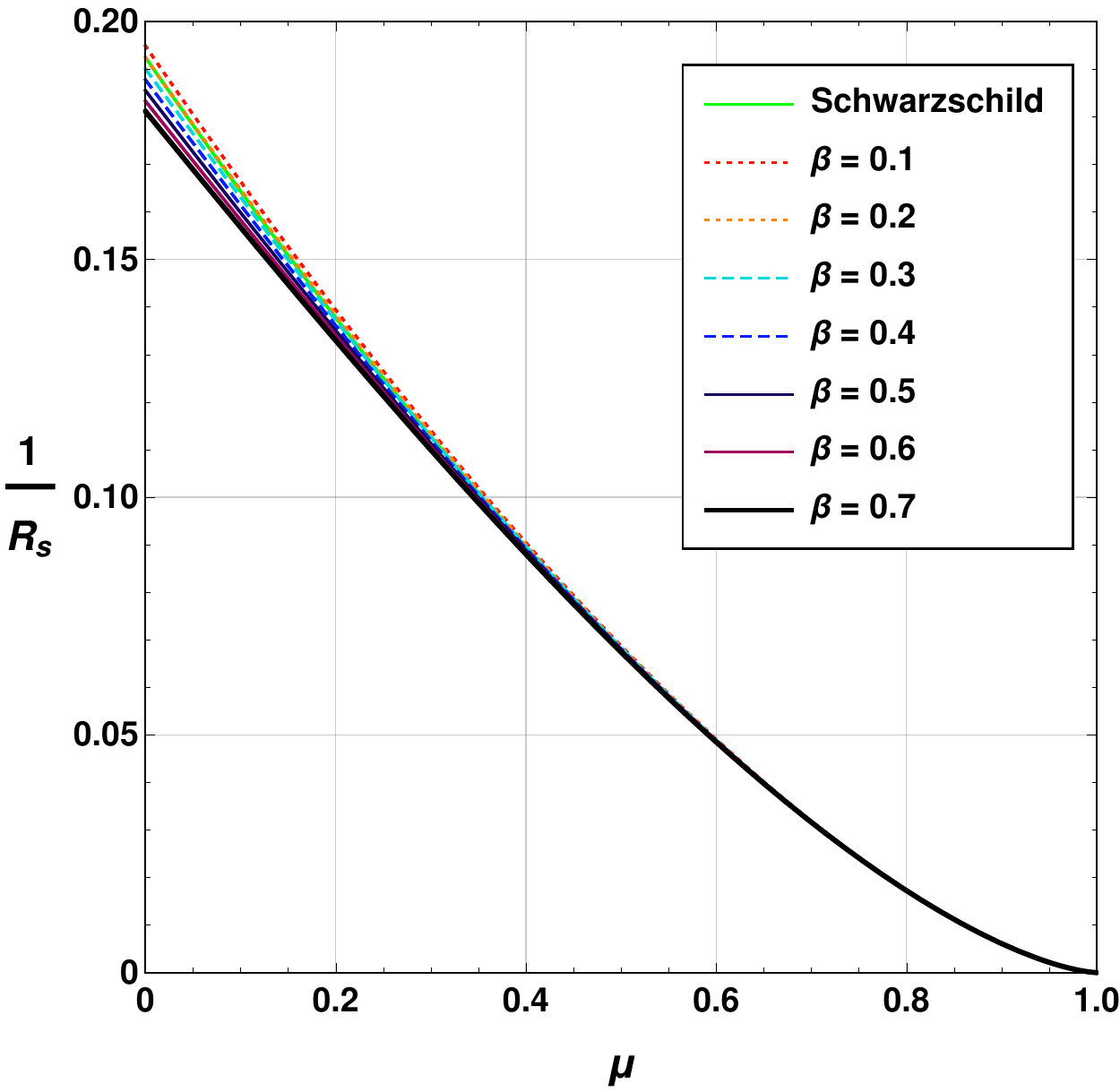}
}
\caption{Variation of $\frac{1}{R_s}$ with respect to $\mu$ for $M=1$. On the 
left panel we use $\beta =0.1$ and on the right panel we use 
$\alpha = 0.1$.}
\label{Shadow03}
\end{figure}

In Fig.\ \ref{Shadow01}, on the left panel, we show the stereographic 
projection of the shadow of the black hole for different values of the black 
hole mass $M$. As usual, with an increase in the mass, the shadow radius of 
the black hole increases. On the right panel, we show the shadow of the 
black hole for different values of the monopole parameter $\mu$. It is seen 
that with an increase in the value of $\mu$, the shadow radius increases in
agreement with the earlier observation from Fig.\ \ref{Shadow00b} of the
black hole potential.
Fig.\ \ref{Shadow02} shows the shadow of the black hole for different values 
of the GUP parameters $\alpha$ and $\beta$. As evident from the analysis of 
the black hole potential, both parameters have opposite impacts on the black 
hole shadow. When $\alpha$ increases, the shadow size decreases gradually. 
However, the impact of $\alpha$ is smaller and opposite compared to $\mu$. 
The other GUP parameter $\beta$ has the smallest impact on the size of the 
black hole shadow as seen from the graph on the right panel of 
Fig.\ \ref{Shadow02}. With an increase in the parameter $\beta$, the shadow 
radius of the black hole increases slowly. To have a more clear visualisation, 
we plot the inverse of the black hole shadow radius with respect to 
the global monopole parameter $\mu$ for different values of the GUP parameters 
in Fig.\ \ref{Shadow03}. On the left panel of Fig.\ \ref{Shadow03}, we  
consider different values of $\alpha$. One can see that $\frac{1}{R_s}$ 
increases with an increase in the values of $\alpha$ for smaller values of 
$\mu$. However, for large values of $\mu$, the GUP parameter has negligible 
impacts on the inverse of the black hole shadow radius. From the right panel 
of Fig.\ \ref{Shadow03}, it is seen that the inverse of the shadow radius 
decreases for smaller values of $\mu$. Again for higher values of 
$\mu$, the impact of GUP parameter $\beta$ is negligible.

\section{A Classical test: Advancement of the perihelion of planets}
\label{sec5}
The geodesic equation for the motion of a test particle along a path described 
by the four-coordinate $x^{\mu}(\tau)$ is given by 
\begin{equation}
\frac{d^2 x^{\mu}}{d\tau^2}+\Gamma^{\mu}_{\sigma\nu} \frac{dx^{\sigma}}{d\tau}\frac{dx^{\nu}}{d\tau}=0,
\label{Eq:20}
\end{equation}
where $\tau$ is the affine parameter. We consider a constant of motion 
$\chi$ associated with the geodesic, which is defined as
\begin{equation}
\chi=-\,g_{\mu\nu}U^{\mu}U^{\nu},
\label{Eq:21}
\end{equation}
where the vector $U^\mu$ is of the form:
\begin{equation}
U^{\mu}=\frac{dx^{\mu}}{d\tau}=\dot{x^{\mu}}.
\label{Eq:22}
\end{equation} 
Here, the differentiation with respect to $\tau$ parameter is represented by 
dot over the variable. For massive particles, $\chi=1$ and for massless 
particles, $\chi=0.$ Using Eq.\ \eqref{Eq:20}, the massless particle's 
trajectory in the spacetime \eqref{Eq:10} can be stated in the form of the 
following equations:
\begin{align}
\ddot{t}-\frac{A}{2 r\Big(A-4 M r (1-\mu)\Big)}\,\dot{r}\,\dot{t}&=0,
\label{Eq:23}\\[8pt]
\ddot{r} +\frac{A}{2 r(A-4 M r(1-\mu))}\, \dot{r}^2 -\frac{A\Big(A -4M r (1-\mu)\Big)}{32\,M^2 r^3 (1+\lambda)}\, \dot{t}^2 \notag\\
+\, \frac{A}{4M(1+\lambda)}\, \dot{\theta}^2 + \frac{A-4M r(1-\mu)}{4M(1+\lambda)} \, \sin^2 \theta \, \dot{\phi}^2&=0,
\label{Eq:24}
\end{align}
\begin{align}
\ddot{\theta}+ \frac{2}{r} \, \dot{r} \, \dot{\theta}- \cos \theta \sin \theta\, \dot{\phi}^2&=0,
\label{Eq:25}\\[8pt]
\ddot{\phi} +\frac{2}{r}\,\dot{r}\,\dot{\phi}+2 \cot \theta\, \dot{\theta}\,\dot{\phi}&=0,
\label{Eq:26}
\end{align}
where $A=8M^2 -2M \alpha (1-\mu)+\beta (1-\mu)^2$. It is to be noted that the 
affine parameter $\tau$ is chosen to be the proper time in case of massive 
test particles and the initial conditions are $\theta(\tau_0)=\frac{\pi}{2}$ 
and $\dot{\theta}(\tau_0)=0$. 
From Eq.\ \eqref{Eq:21}, for the massive test particle in timelike geodesic, 
we have the following differential equation for the coordinate $r$:
\begin{equation}
(1+\lambda)\,\dot{r}^2 +\Big(1+\frac{L^2}{r^2}\Big)\Big[1-\mu-\frac{2M}{r}\Big(1-\frac{\alpha(1-\mu)}{4M}+\frac{\beta(1-\mu)^2}{8M^2}\Big)\Big]=E^2.
\label{Eq:29}
\end{equation}                                                                  
Following Ref.\ \cite{17}, we introduce a new variable $m=r^{-1}$ such that 
we have 
\begin{equation}
\dot{r}=-\,L\,\frac{dm}{d\phi}.
\label{Eq:30}
\end{equation}
By using Eq.\ \eqref{Eq:30} in Eq.\ \eqref{Eq:29}, we can write
\begin{equation}
m^2(1-\mu )+(1+\lambda)\left(\frac{dm}{d\phi}\right)^2=\frac{E^2-(1-\mu )}{L^2}+2M m\left(\frac{1}{L^2}+m^2\right)\bigg[1-\frac{\alpha(1-\mu)}{4M}+\frac{\beta(1-\mu)^2}{8M^2}\bigg].
\label{Eq:31}
\end{equation}
Taking the derivative of this Eq.\ \eqref{Eq:31} with respect to $\phi$, we 
obtain 
\begin{equation}
(1+\lambda)\,\frac{d^2 m}{d\phi^2}+m(1-\mu)-\bigg(\frac{M}{L^2}+3\,m^2 M\bigg)\bigg[1-\frac{\alpha(1-\mu)}{4M}+\frac{\beta(1-\mu)^2}{8M^2}\bigg]=0.
\label{Eq:32}
\end{equation}

The above differential equation in one variable contains contributions of both 
Lorentz symmetry violation parameter and global monopole term. In order to 
solve this type of equation, we use the perturbative method, for which define 
the variable $m$ as
\begin{equation}
m=m^{(0)}+\epsilon\, m^{(1)},
\label{Eq:33}
\end{equation} 
where the small perturbative parameter $\epsilon=\frac{3M^2}{L^2}$ and 
$\epsilon<<1$ \cite{17}. Using this, the Eq.\ \eqref{Eq:32} for the zeroth 
order in $\epsilon$ gives
\begin{equation}
(1+\lambda)\frac{d^2 m^{(0)}}{d\phi^2}+m^{(0)}(1-\mu)-\frac{M}{L^2}\bigg[1-\frac{\alpha(1-\mu)}{4M}+\frac{\beta(1-\mu)^2}{8M^2}\bigg]=0.
\label{Eq:34}
\end{equation} 
Solution of this differential Eq.\ \eqref{Eq:34} for zeroth order in 
$\epsilon$ gives 

\begin{equation}
 m^{(0)}=\frac{M_{GUP}}{(1-\mu)L^2}\bigg[1+e \cos \frac{\phi \sqrt{1-\mu}}{\sqrt{1+\lambda}}\bigg].
\label{Eq:35}
\end{equation}

It is to be noted that when we substitute $\alpha=\beta=\mu=0$ in the above 
Eq.\ \eqref{Eq:35}, we get back the expected GR result. Here $e$ is the 
eccentricity of the orbit. We then implement the differential 
Eq.\ \eqref{Eq:32} for the variable in Eq.\ \eqref{Eq:33} to the first order 
in $\epsilon$ to get the differential equation as
\begin{equation}
 (1+\lambda)\frac{d^2 m^{(1)}}{d\phi^2}+m^{(1)}(1-\mu)-\frac{L^2 m^{(0)2}}{M_{GUP}}=0.
\label{Eq:36}
\end{equation}
Solutions of the two differential Eqs.\ \eqref{Eq:34} and \eqref{Eq:36} when 
clubbed in Eq.\ \eqref{Eq:33} gives an ellipse-like equation after the small 
$\epsilon$ approximation as

\begin{equation}
 m=\frac{M_{GUP}}{(1-\mu)L^2}\bigg[1+e\Big(\cos \frac{\sqrt{1-\mu}}{\sqrt{1+\lambda}}\phi (1-\epsilon)\Big)\bigg].
\label{Eq:37}
\end{equation}
Eq.\ \eqref{Eq:37} represents the perturbative solution of the original 
differential Eq.\ \eqref{Eq:32}. It is quite clear from the expression of 
period that the GUP parameters play no role regarding the precession of orbits. Despite the 
presence of different factors of model considered in the study, the orbit is 
periodic as can be seen from the expression with a period,
\begin{equation}
\Phi=\frac{2\pi \sqrt{1+\lambda}}{\sqrt{1-\mu}(1-\epsilon)}=2\pi+\Delta \Phi.
\label{Eq:38}
\end{equation}
The quantity $\Delta \Phi$ represents the advance of the perihelion of the 
massive object (planet) which is expressed by expanding the above expression 
in lowest order of $\epsilon$, $\lambda$ and $\mu$ as
\begin{equation}
\Delta \Phi=2\pi \epsilon+ \pi \lambda+\pi \mu +\frac{\pi \lambda\mu}{2} =\Delta \Phi_{GR} +\Delta\Phi_{LV}+\Delta \Phi_{GM}+\Delta \Phi_{LVGM}.
\label{Eq:39}
\end{equation}
 Here, $\Delta \Phi_{GR}$, $\Delta\Phi_{LV}$, $\Delta \Phi_{GM}$ and $\Delta \Phi_{LVGM}$ means usual GR contribution, 
contribution due to Lorentz violation, contribution due to global monopole 
term, and a mixed contribution from both the Lorentz violation and global 
monopole respectively. 
It is clear that along with $2\pi \epsilon$ term which appears in GR, we have 
some extra contributions to the period of the orbit. 

The perihelion precession data of various planets and an asteroid Icarus whose 
time periods of revolution around the sun in elliptical orbits are known from 
various experimental observations \cite{17}. The uncertainty associated with 
these precession data is thus available which is assumed to be the upper 
bounds of any deviation from GR. But the issue which we now face is that 
there are two quantities $\lambda$ and $\mu$ in our case, which makes the 
exact determination hard from one single constraint equation. Thus we can 
estimate that the product of these two quantities is less than this bound as 
will be clear from the table \ref{Table01} below as well as from 
Eq.\ \eqref{Eq:39}.

Here we present a tabular data of precession of some inner planets and an 
asteroid in units of arc-seconds per century and compare the GR-predicted and 
observed values and also present an estimate of the upper bounds on the 
parameters $\lambda$ and $\mu$. Stringent upper bounds are obtained from the 
precession data of Earth and Mars of the order of $10^{-12}$, while for the 
asteroid Icarus, the upper bound obtained is of the order $10^{-8}$ only.
\begin{table}[h!]
\centering
\caption{The precession data, the orbital periods and upper bounds of model
parameters for some planets and the asteroid Icarus \cite{17,57,58,59,60,61}. 
Precession data is in the unit of arc-second per century.}
\begin{tabular}{ccccc}
\\
\hline \hline
Planet/Asteroid & Predicted by GR & Observation with error & Orbital period (in days)  & Upper Bounds for $(\lambda +\mu + \frac{\lambda \mu }{2})$\\
\hline
 
Mercury  & $42.981$ & $42.981 \pm 0.0030$ &  $88.0$ & $1.1 \times 10^{-11}$\\

 Venus  & $8.6247$ & $8.6273\pm 0.0016$ & $224.7$ & $1.5 \times 10^{-11}$\\

 Earth  & $3.83877$ & $3.83896\pm 0.00019$ & $365.2$ & $2.9 \times 10^{-12}$\\

 Mars  & $1.350938$ & $1.350918\pm 0.000037$ & $687.0$ & $1.1 \times 10^{-12}$\\

 Icarus (asteroid) & $10.10$ & $9.80\pm 0.80$ & $409.0$ & $1.3 \times 10^{-8}$ \\
 
 \hline \hline
\end{tabular}
\label{Table01}
\end{table}

It is to be noted that the bounds are similar to the ones obtained in 
\cite{17}, where the authors employed the LSB only. However, we obtained 
additional bounds on $\mu$, the global monopole parameter and the product of 
$\mu$ an $\lambda$ as shown in the table \ref{Table01}. Another point that has 
to be mentioned is that the perihelion precession test of planetary orbits 
is very well explained by GR theory and thus very small deviations in the 
form of errors are seen. This is the reason for the very minute values of the 
parameters obtained as bounds.

\section{Conclusion} \label{conclusion}
\label{sec6}
In this work we have studied the various thermodynamic relations and
shadows for a spherically symmetric and static GUP-corrected 
Schwarzschild-type black hole solution which contains the topological defects 
and Lorentz symmetry violation in Bumblebee gravity, presented in the 
Ref.\ \cite{20}. The black hole temperature variation with respect to
black hole mass for various parameters of the model was almost identical in 
the sense that initially the temperature increases to a peak and then it 
decreases continuously with increase in the mass of the black hole. It 
may be highlighted that as the black hole increases in mass, its temperature 
gradually decreases and approaches zero. But it does not turn negative 
and hence there is no possibility of the formation of an ultracold black 
hole \cite{62}. Thermodynamically stable black holes have positive 
heat capacity and the one we studied did not show this property. 
We studied the variation of heat capacity with various parameters 
of our theory. It is to be noted that when a black hole absorbs more 
energy than it is throwing out, its mass will increase indefinitely, 
whereas when it emits more than it absorbs, then it will eventually 
disappear. Such is the situation with negative heat capacities, 
wherein emission is more than absorption, causing instability. 
In this regard, our black hole spacetime shows instability similar to 
Schwarzschild case. We study the entropy function and its variation with 
respect to mass for various values of our model parameters. It was found that 
entropy increases with mass for all the cases. Gibbs free energy variations 
have been studied and we see similar a increasing trend in variation for all 
four model parameters.

We studied the shadow radius for the black hole with variations of different 
parameters of the theory. The shadow of the black hole is a suitable optical 
characteristic which plays a significant role from the observational 
perspective. Our investigation shows that the presence of the global monopole 
has a significant impact on the shadow radius of the black hole. An 
interesting fact is that when the global monopole parameter increases, the 
effects of the GUP parameters become negligible. It basically implies that it 
might be challenging to probe the quantum corrections like GUP corrections 
associated with a black hole in presence of a global monopole in spacetime 
by utilising the shadow analysis of the black hole. Another result obtained 
here is that the global monopole and the second GUP parameter $\beta$ have 
similar impacts on the black hole shadow. Although the peak of the reduced 
potential of the black hole is significantly affected by the Lorentz symmetry 
breaking, one may note that photon sphere position is independent of the 
Lorentz symmetry violation. As a result, we do not have any impacts on the 
shadow radius by the Lorentz symmetry violation. In recent years, the Event 
Horizon Telescope (EHT) has made significant strides in its efforts to capture 
an ultra-high resolution image of the accretion flows surrounding a 
supermassive black hole in the galaxy M87$^*$. These efforts have finally 
culminated in the acquisition of a groundbreaking image that showcases 
the inner workings of the black hole's accretion process \cite{6,7,8,9,10,11}.

The first image of M87$^*$ reveals a strikingly bright ring encircling the 
black hole's dark interior. This ring, known as the photon ring, serves as a 
critical observation feature of the black hole. Meanwhile, the black hole's 
dark center, 
known as its shadow, is prominently displayed in the image. These observations 
are groundbreaking in their ability to shed light on the mysterious nature of 
supermassive black holes and the processes that govern their behavior. An 
extension of this current work may be constraining the shadow radius using 
the observational results.  In view of observational data related to 
shadow radius made available by EHT recently, it is possible to put 
constraints on various parameters like GUP and the monopole parameter. This 
remains as a future extension of our work.

We did a classical test of GR also, namely the perihelion precession of inner 
planets and an asteroid. It is seen that the GUP parameters do not 
contribute towards the period of precession of the planetary orbits. 
Moreover, the upper 
bounds obtained are not very stringent as number of parameters is two in our 
case. It can be concluded that this method of constraining the parameters 
works best when we work with one parameter as in case of \cite{17}. 
Various other tests like bending of light and Shapiro time 
delay of light have been used to constrain the parameters of the theory, which 
we have not dealt with now and keep it as a future scope of study.



\bibliographystyle{apsrev}

\end{document}